\documentclass[apj]{emulateapj}
\usepackage{epsf}
\usepackage{amsmath,graphicx} 
\usepackage[z,alpha]{tmsfnts}
\def\Mgas{\ensuremath{M_{\rm g}}}

\def\Mg{\ensuremath{M_{\rm g}}}

\def\TmgSIM{\ensuremath{T_{\text{mg}}^{\text{SIM}}}}

\def\Tx{\ensuremath{T_{\text{X}}}}
\def\Lx{\ensuremath{L_{\text{X}}}}
\def\Yx{\ensuremath{Y_{\text{X}}}}

\def\rhocrit{\ensuremath{\rho_{\text{crit}}}}

\def\Msun{\ensuremath{M_\odot}}

\def\r500{\ensuremath{r_{\text{500}}}}
\def\M500{\ensuremath{M_{\text{500}}}}
\def\T500{\ensuremath{T_{\text{500}}}}
\def\P500{\ensuremath{P_{\text{500}}}}
\def\K500{\ensuremath{K_{\text{500}}}}

\def\***#1{\textbf{\textsf{***#1***}}}

\def\nls{\\[-5pt]}
\def\nlss{\\[-2pt]}

\shorttitle{EFFECTS OF GALAXY FORMATION ON ICM} 
\shortauthors{NAGAI, KRAVTSOV, \& VIKHLININ}

\begin{document}

\slugcomment{The Astrophysical Journal, 668:1-14, 2007 October 10}

\title{Effects of Galaxy Formation on Thermodynamics of the Intracluster
  Medium} \author{Daisuke Nagai\altaffilmark{1}, Andrey V.
  Kravtsov\altaffilmark{2}, Alexey Vikhlinin\altaffilmark{3,4}}

\begin{abstract} 
  We present detailed comparisons of the intracluster medium (ICM) in
  cosmological Eulerian cluster simulations with deep \emph{Chandra}
  observations of nearby relaxed clusters.  To assess the impact of
  galaxy formation, we compare two sets of simulations, one performed
  in the non-radiative regime and another with radiative cooling and
  several physical processes critical to various aspects of galaxy
  formation: star formation, metal enrichment and stellar feedback. We
  show that the observed ICM properties outside cluster cores are
  well-reproduced in the simulations that include cooling and star
  formation, while the non-radiative simulations predict an overall
  shape of the ICM profiles inconsistent with observations. In
  particular, we find that the ICM entropy in our runs with cooling is
  enhanced to the observed levels at radii as large as half of the
  virial radius.  We also find that outside cluster cores entropy
  scaling with the mean ICM temperature in both simulations and
  \emph{Chandra} observations is consistent with being self-similar
  within current error bars.  We find that the pressure profiles of
  simulated clusters are also close to self-similar and exhibit little
  cluster-to-cluster scatter. We provide analytic fitting formulae for
  the pressure profiles of the simulated and observed clusters.  The
  X-ray observable-mass relations for our simulated sample agree with
  the \emph{Chandra} measurements to $\approx 10\%-20\%$ in
  normalization.  We show that this systematic difference could be
  caused by the subsonic gas motions, unaccounted for in X-ray
  hydrostatic mass estimates.  The much improved agreement of
  simulations and observations in the ICM profiles and scaling
  relations is encouraging and the existence of tight relations of
  X-ray observables, such as $Y_X$, and total cluster mass and the
  simple redshift evolution of these relations hold promise for the
  use of clusters as cosmological probes. However, the disagreement
  between the predicted and observed fractions of cluster baryons in
  stars remains a major puzzle.
\end{abstract}


\altaffiltext{1}{Theoretical Astrophysics, California Institute of
Technology, Mail Code 130-33, Pasadena, CA 91125 ({\tt daisuke@caltech.edu})}
\altaffiltext{2}{Department of Astronomy and
Astrophysics, Kavli Institute for Cosmological Physics, Enrico Fermi Institute, 5640 South
Ellis Ave., University of Chicago, Chicago, IL 60637}
\altaffiltext{3}{Harvard-Smithsonian Center for Astrophysics, 60 Garden Street, Cambridge, MA 02138}
\altaffiltext{4}{Space Research Institute, 84/32 Profsojuznaya Street, GSP-7, Moscow 117997, Russia}

\section{Introduction}
\label{sec:intro}

Clusters of galaxies are fascinating astrophysical objects and
laboratories for studying galaxy formation and structure formation in
general.  At the same time, clusters can provide cosmological
constraints that are complementary to those obtained with other
methods such as temperature anisotropies of the cosmic microwave
background, Type Ia supernovae, and weak lensing
\citep[e.g.,][]{voit05,tozzi06,borgani06,albrecht_etal06}. Cosmological
applications of clusters include cluster counts and their evolution
with redshift
\citep[e.g.,][]{henry_arnaud91,markevitch98,ikebe_etal02,vikhlinin_etal03},
spatial distribution \citep[e.g.,][]{miller_etal02}, and the
angular-diameter distance measurements
\citep{allen_etal04,laroque_etal06}. Detailed observations of merging
clusters provide unique insights into the physics of the intracluster
plasma
\citep[e.g.,][]{vikhlinin_etal01a,vikhlinin_etal01b,markevitch_etal03}
and provide key evidence for the existence and properties of dark
matter \citep[][]{markevitch_etal04,clowe_etal06}.

All cosmological applications of clusters, at least to a certain
degree, rely on solid understanding of the physics of their
formation. Given that clusters are nonlinear collapsed systems,
numerical cosmological simulations are the method of choice for their
theoretical studies.  Modern cosmological codes using $N$-body and
numerical hydrodynamics techniques can accurately follow dynamics of
dark matter and gaseous baryonic components in their full complexity
during the hierarchical build-up of structures.  Yet, more realistic
modeling of clusters requires inclusion of additional baryonic
processes. For example, to model formation of cluster galaxies, we
need, at the very least, to correctly treat energy dissipation due to
radiative losses by baryons, and conversion of gas into stars. In
addition, any feedback in the form of energy injection and metal
enrichment from supernova winds
\citep[e.g.,][]{metzler_evrard94,valdarnini03} and active galactic
nuclei \citep{brueggen_etal05,sijacki_springel06,cattaneo_teyssier07},
and injection of non-thermal cosmic rays at large-scale shocks
accompanying cluster formation \citep{pfrommer_etal07} can alter the
thermodynamics of the intracluster gas.

Although our understanding of details and relative importance of these
processes is currently sketchy, the simulations with specific
assumptions about them are highly predictive, which should make models
falsifiable.  In particular, by comparing observed cluster properties
with the results of simulations that include various physical processes
described above we can learn a great deal about these processes and
their role in cluster formation.

Over the last two decades, such comparisons were used extensively to
put constraints on the deviations of ICM thermodynamics from the
simple self-similar behavior, described originally by
\citet{kaiser86,kaiser91}.  The first studies of the observed
correlation of cluster X-ray luminosity, $\Lx$, and spectral
temperature, $\Tx$, unambiguously showed that its slope is steeper
than the slope predicted by the self-similar model
\citep[e.g.,][]{edge_stewart91,henry_arnaud91,white_etal97,markevitch98,allen_fabian98,arnaud_evrard99}.
In addition, the slope of the $\Lx-\Tx$ relation steepens for the
lowest mass clusters
\citep[e.g.,][]{helsdon_ponman00,finoguenov_etal02,finoguenov_etal07}.
Deviations from self-similarity were shown to be the strongest in the
cores of clusters
\citep[e.g.,][]{markevitch98,degrandi_molendi02,vikhlinin_etal06} and
were widely interpreted as evidence for preheating of the intracluster
gas by energy from supernovae and AGN feedback
\citep[e.g.,][]{david_etal91,kaiser91,evrard_henry91,white91,wu_etal00,bialek_etal01,borgani_etal01,borgani_etal02,nath_etal02}.

Alternative explanation was proposed by \citet{bryan00}, who argued
that cooling and condensation of the gas accompanying formation of
cluster galaxies can reduce the ICM gas density and increase its
temperature and entropy to the observed levels \citep[see
also][]{voit_bryan01,voit_etal02}. This explanation was borne out by
cosmological simulations
\citep[][]{pearce_etal00,muanwong_etal01,valdarnini02,dave_etal02,kay_etal04,kay_etal07}.
However, the amount of gas that condenses out of the hot ICM in
cosmological simulations due to cooling
\citep[e.g.,][]{suginohara_ostriker98,lewis_etal00,pearce_etal00,dave_etal02,ettori_etal04}
appears to be a factor of 2-3 too large compared to the observed
stellar mass in clusters \citep[][]{lin_etal03,gonzalez_etal07}.
Thus, the X-ray measurements appear to be consistent with a large
fraction of cooling gas, while the optical estimates of stellar mass
indicate that this fraction is small.

Modern X-ray observations with \emph{Chandra} and \emph{XMM-Newton}
allow us to study the ICM properties with unprecedented detail and
accuracy. Their superb spatial resolution and sensitivity enable
resolved, accurate X-ray brightness and temperature maps over a large
fraction of the cluster virial radii. The X-ray measurements also
enable accurate mass modeling of relaxed clusters with the assumption
of hydrostatic equilibrium of the ICM in the cluster potential. These
observations can therefore be used for detailed comparisons of both
global cluster properties and their profiles with simulation results,
which provide more stringent tests for the models of the ICM
thermodynamics.  In particular, such comparisons can shed some light
on the apparently contradictory lines of evidence as to the efficiency
of cooling in clusters described above. 

To this end, in the present study we focus on the effects of radiative
cooling and star formation on the observable properties of clusters
and compare results of simulations in both non-radiative and radiative
regimes with the current X-ray data. Namely, we use two sets of
simulations started from the same initial conditions. Both sets treat
collisionless dynamics of dark matter and hydrodynamics of diffuse gas
with high-resolution using the adaptive mesh refinement technique. In
the baseline set of cluster simulations, the gas is modeled in
non-radiative regime and thus does not reach high densities and is not
allowed to form stars. The second set of simulations includes several
processes accompanying galaxy formation: gas cooling, star formation,
metal enrichment and thermal feedback due to the supernovae.
Comparison of the simulated profiles in these two sets of simulations
to those of observed clusters allows us to gauge the role of galaxy
formation in shaping properties of the ICM.  As we show in
\S~\ref{sec:profiles}, the simulations that include galaxy formation
processes provide a considerably better match to the observed ICM
profiles outside cluster cores compared to the non-radiative
simulations.

The paper is organized as follows. In \S~\ref{sec:cosmo} we describe
cosmological cluster simulations. The methods used to analyze the
simulations and the brief description of observations used in our
comparisons are given in \S~\ref{sec:cosmo} and \ref{sec:obs},
respectively. We present results of comparison of the ICM density,
temperature, entropy, and pressure profiles in \S~\ref{sec:profiles},
and integrated quantities such as spectral X-ray gas temperature, gas
mass, and pressure, in simulations and observations in
\S~\ref{sec:obs_comp}.  We discuss our results and conclusions in
\S~\ref{sec:discussion}. 

\section{Cosmological Cluster Simulations}
\label{sec:cosmo}

In this study, we analyze high-resolution cosmological simulations of
16 cluster-sized systems in the flat {$\Lambda$}CDM model:
$\Omega_{\rm m}=1-\Omega_{\Lambda}=0.3$, $\Omega_{\rm b}=0.04286$,
$h=0.7$, and $\sigma_8=0.9$, where the Hubble constant is defined as
$100h{\ \rm km\ s^{-1}\ Mpc^{-1}}$, and an $\sigma_8$ is the power
spectrum normalization on an $8h^{-1}$~Mpc scale.  The simulations
were done with the Adaptive Refinement Tree (ART)
$N$-body$+$gasdynamics code \citep{kravtsov99, kravtsov_etal02}, an
Eulerian code that uses adaptive refinement in space and time, and
(non-adaptive) refinement in mass \citep{klypin_etal01} to reach the
high dynamic range required to resolve cores of halos formed in
self-consistent cosmological simulations. The same set of cluster
simulations was used in our related recent studies
\citep{kravtsov_etal06,nagai_etal07}, where additional details can be
found. We provide a description of the simulation details here for
completeness.

%
%
%
%
%
%

\begin{deluxetable*}{lccccccccc}
  \tablecolumns{8} \tablecaption{Simulated Cluster Sample of the CSF Run at
    $z$=0\label{tab:sim}} \tablehead{ \multicolumn{1}{c}{Name\hspace*{7mm}}&
    \multicolumn{1}{c}{}& \multicolumn{1}{c}{$\r500$} &
    \multicolumn{1}{c}{$\M500^{\rm gas}$} & \multicolumn{1}{c}{$\M500^{\rm
        tot}$} & \multicolumn{1}{c}{$\TmgSIM$} & \multicolumn{3}{c}{$\Tx$} & 
        \multicolumn{1}{c}{rel.$/$unrel.} 
    \\
    \multicolumn{2}{c}{}& \multicolumn{1}{c}{($h^{-1}$Mpc)}&
    \multicolumn{2}{c}{($h^{-1}10^{13}\Msun$)}&
    \multicolumn{1}{c}{(keV)$\tablenotemark{a}$}&
    \multicolumn{3}{c}{(keV)$\tablenotemark{b}$}&
    \multicolumn{1}{c}{(1/0)$\tablenotemark{c}$} } 
  \startdata
  CL101 \dotfill & & 1.160 & 8.17 & 90.81 & 7.44 & 8.72 & 8.67 & 8.86 & 000 \\
  CL102 \dotfill & & 0.978 & 4.82 & 54.47 & 5.63 & 5.63 & 5.83 & 5.86 & 000 \\
  CL103 \dotfill & & 0.994 & 4.92 & 57.71 & 4.84 & 4.73 & 4.93 & 4.62 & 000 \\
  CL104 \dotfill & & 0.976 & 5.15 & 53.88 & 6.61 & 7.69 & 7.73 & 7.73 & 111 \\
  CL105 \dotfill & & 0.943 & 4.71 & 48.59 & 5.67 & 6.21 & 6.21 & 6.17 & 001 \\
  CL106 \dotfill & & 0.842 & 3.17 & 34.65 & 4.54 & 4.34 & 4.35 & 4.30 & 000 \\
  CL107 \dotfill & & 0.762 & 2.17 & 25.66 & 3.61 & 3.97 & 3.71 & 3.94 & 100 \\
  CL3   \dotfill & & 0.711 & 1.91 & 20.90 & 3.37 & 3.65 & 3.60 & 3.61 & 111 \\
  CL5   \dotfill & & 0.609 & 1.06 & 13.11 & 2.22 & 2.40 & 2.39 & 2.39 & 111 \\
  CL6   \dotfill & & 0.661 & 1.38 & 16.82 & 2.88 & 3.38 & 3.38 & 3.57 & 000 \\
  CL7   \dotfill & & 0.624 & 1.21 & 14.13 & 2.54 & 2.96 & 2.88 & 2.90 & 111 \\
  CL9   \dotfill & & 0.522 & 0.73 & 8.23  & 1.58 & 1.53 & 1.60 & 1.57 & 000 \\
  CL10  \dotfill & & 0.487 & 0.43 & 6.72  & 1.58 & 1.93 & 1.90 & 1.91 & 111 \\
  CL11  \dotfill & & 0.537 & 0.78 & 8.99  & 1.75 & 2.00 & 2.02 & 1.98 & 000 \\
  CL14  \dotfill & & 0.509 & 0.62 & 7.69  & 1.64 & 1.85 & 1.84 & 1.83 & 111 \\
  CL24  \dotfill & & 0.391 & 0.26 & 3.47  & 0.97 & 1.06 & 1.04 & 1.07 & 010 \enddata
  \tablenotetext{a}{$\TmgSIM$ is the average temperature measured
    directly from the 3D ICM distribution in the simulations.} 
  \tablenotetext{b}{Average temperatures measured in the shell of
    [0.15,1]$\r500$ from the mock \emph{Chandra} analysis of simulated
    clusters viewed along three orthogonal projection axes 
    ($x$, $y$, $z$, {\it from left to right}). Note that the values 
of $\Tx$ quoted here are different from those in Table 1 of \citet{nagai_etal07}, where
erroneous values were presented.} 
  \tablenotetext{c}{Classification of relaxed and unrelaxed clusters are indicated with 0
  and 1, respectively, for the three projections.} 
\end{deluxetable*}

The $N-$body$+$gasdynamics cluster simulations used in this analysis
follow collisionless dynamics of dark matter and stars, gasdynamics
and several physical processes critical to various aspects of galaxy
formation: star formation, metal enrichment and thermal feedback due
to Type II and Type Ia supernovae, self-consistent advection of
metals, metallicity-dependent radiative cooling and UV heating due to
cosmological ionizing background \citep{haardt_madau96}.  The cooling
and heating rates take into account Compton heating and cooling of
plasma, UV heating, and atomic and molecular cooling, and are
tabulated for the temperature range $10^2<T<10^9$~K and a grid of
metallicities, and UV intensities using the Cloudy code \citep[ver.
96b4;][]{ferland_etal98}. The Cloudy cooling and heating rates take
into account metallicity of the gas, which is calculated
self-consistently in the simulation, so that the local cooling rates
depend on the local metallicity of the gas.  Star formation in these
simulations was done using the observationally-motivated recipe
\citep[e.g.,][]{kennicutt98}: $\dot{\rho}_{\ast}=\rho_{\rm
gas}^{1.5}/t_{\ast}$, with $t_{\ast}=4\times 10^9$~yrs. Stars are
allowed to form in regions with temperature $T<2\times10^4$K and gas
density $n > 0.1\ \rm cm^{-3}$.
\footnote{We have compared runs where star formation was allowed to
proceed in regions different from our fiducial runs. We considered
thresholds for star formation of $n=10$, $1$, $0.1$, and $0.01\ {\rm
cm^{-3}}$. We find that thresholds affect the properties of the ICM at
small radii, $r/r_{\rm vir}<0.1$, but differences are negligible at the radii we
consider in this study.}  The code also accounts for the stellar
feedback on the surrounding gas, including injection of energy and
heavy elements (metals) via stellar winds, supernovae, and secular
mass loss.  The details of star formation prescription and feedback
are discussed in \citet{kravtsov_etal05}. Some potentially relevant
physical processes, such as AGN bubbles, physical viscosity, magnetic
field, and cosmic rays, are not included. 

The adaptive mesh refinement technique is used to achieve high spatial
resolution in order to follow the galaxy formation and evolution
self-consistently in these simulations.  The peak spatial resolution
is $\approx 7$ and $5\,h^{-1}$~kpc, and the dark matter particle mass
in the region around the cluster was $9.1\times 10^{8}$ and $2.7\times
10^{8}\,h^{-1}\,M_{\odot}$ for CL 101--107 and CL 3--24, respectively.
To test the effects of galaxy formation, we also repeated each cluster
simulation with only the standard gasdynamics for the baryonic
component, without radiative cooling or star formation.  We will use
labels ``non-radiative'' and ``cooling+SF'' (CSF) to refer to these
two sets of runs, respectively.

In this work, we also use mock \emph{Chandra} X-ray images and spectra
of the simulated clusters to derive total mass, gas mass and
temperature profiles, as well as integrated cluster properties, using
the analysis procedures essentially identical to those used to analyze
real \emph{Chandra} observations, as described in
\citet{nagai_etal07}. The average X-ray spectral temperature, $\Tx$,
is obtained from a single-temperature fit to the spectrum integrated
within $r_{500}$, excluding the central region, $r<0.15r_{500}$.  For
each cluster, the mock data is created for three orthogonal
projections along the $x$, $y$, and $z$ coordinate axes. In
\S~\ref{sec:scaling} we use quantities derived from the mock
observations to compare scaling relations exhibited by simulated
clusters to observations. 

Our simulated sample includes 16 clusters at $z=0$ and their most
massive progenitors at $z=0.6$. The properties of simulated clusters
at $z=0$ are given in Table~\ref{tab:sim}. The masses are reported at
the radius $\r500$ enclosing overdensities with respect to the
critical density at the redshift of the output.  This choice of the
outer radius is mainly motivated by the fact that clusters are more
relaxed within $r_{500}$ compared to the outer regions
\citep{evrard_etal96}. We also use $r_{200}$, $r_{1000}$, and
$r_{2500}$ which are approximately $1.52$, $0.71$, and $0.44$ times
$r_{500}$, respectively. Mean spectral temperatures are presented
separately for the three orthogonal projections to show the variation
due to projection effects, substructure, etc.  Note that the values of
$\Tx$ quoted here are different from those in Table 1 of
\citet{nagai_etal07}, where erroneous values were presented by
mistake.  In our analysis below we distinguish unrelaxed and relaxed
clusters for a more consistent comparison with the observations. The
classification is based on the overall morphology of the mock X-ray
images, as discussed in \cite{nagai_etal07}.  In Table~\ref{tab:sim}
relaxed and unrelaxed clusters are indicated with 0 and 1 for the
three orthogonal projections ($x$, $y$, $z$ from left to right).

\section{Observational Cluster Sample}
\label{sec:obs}

To test our simulation results against observations we use a set of
accurate measurements of gas density, temperature, and total mass
profiles for a sub-sample of 13 relaxed clusters at $z\sim 0$ that was
presented in \citet{vikhlinin_etal05c} and \citet{vikhlinin_etal06}.
The clusters are selected on the basis of regular and relaxed
morphology of their X-ray surface brightness images, although some of
the systems show signs of AGN activity in their cores.  Three of the
low-$\Tx$ clusters, including USGC~S152 ($\Tx=0.69$~keV), A262
($\Tx=1.89$~keV), and RXJ1159+5531 ($\Tx=1.80$~keV), are excluded from
the comparisons that involve measurements of $M_{500}$ or
normalization with $r_{500}$ because their values are very uncertain
due to insufficient spatial coverage. In \S~\ref{sec:entropy_comp}, we
include USGC~S152 and RXJ1159+5531 for comparisons of the entropy
scaling relations at $0.1r_{200}$, $r_{2500}$, and $r_{1000}$, and
A262 at the first two radii, but not at $r_{1000}$. Since none of the
measurements extends out to $r_{200}$, we estimate $r_{200}$ using
$r_{200}=1.52 r_{500}$, which provides a robust and accurate estimate
of $r_{200}$ for our CSF and non-radiative simulations as well as the
\emph{XMM-Newton} mass measurements \citet{pointecouteau_etal05}. The
observations and analysis procedure used to extract ICM properties and
profiles from the \emph{Chandra} data are described in detail in
\citet{vikhlinin_etal05c} and \citet{vikhlinin_etal06b}.

In our previous study \citep{nagai_etal07}, we used the mock
\emph{Chandra} images and spectra of the simulated clusters to assess
the accuracy of the X-ray measurements of galaxy cluster properties.
Our results show that the X-ray analysis of \citet{vikhlinin_etal06}
provides very accurate reconstruction of the 3D gas density and
temperature profiles for relaxed clusters.  Therefore, we directly
compare the profiles derived from \emph{Chandra} analysis to the 3D
profiles of simulated clusters.  Note that masses and overdensity
radii of the clusters in the \emph{Chandra} sample were derived from
the X-ray hydrostatic analysis. A bias in the estimated cluster mass
may results in a slight underestimate of the estimated cluster virial
radius $r_{500}^{\rm est}$ by about a few percent for relaxed clusters
\citep[see also][for more details and discussions]{nagai_etal07}. We
will show a such comparison in \S~\ref{sec:profiles}.

At the same time, our tests show that X-ray analysis can result in a
$\approx 15\%$ underestimate in the hydrostatic estimates of total
cluster mass. The bias is due to the non-thermal pressure support from
the sub-sonic turbulent motions of the ICM gas, ubiquitous in cluster
simulations
\citep{evrard_etal90,norman_bryan99,nagai_etal03,rasia_etal04,kay_etal04,faltenbacher_etal05,dolag_etal05,rasia_etal06,
nagai_etal07}, but not included in observational hydrostatic mass
estimates.  In \S~\ref{sec:scaling}, we present the comparisons of the
X-ray observable-mass relations of our simulated clusters to deep
\emph{Chandra} X-ray observations of nearby, relaxed clusters using
both the true masses of clusters measured in simulations and the
masses estimated from hydrostatic equilibrium analysis.  We also
correct for the differences in the assumed cosmological parameters in
simulations ($f_b \equiv \Omega_B/\Omega_M = 0.1429$ and $h=0.7$) to
those assumed in the observational analyses ($f_b = 0.175$ and
$h=0.72$).  We adopt $f_b = 0.175$ and $h=0.7$ throughout this work.
Note that $\Omega_M=0.3$ and $\Omega_{\Lambda}=0.7$ are assumed in
both analyses.

In addition, we compare in \S~\ref{sec:entropy_comp} the
\emph{Chandra} ICM entropy measurements with those based on the
\emph{XMM-Newton} observations of 10 clusters \citep{pratt_etal06} and
the \emph{ROSAT+ASCA} data for 64 clusters \citep{ponman_etal03}.  In
these comparisons, we do not use measurements that involve
extrapolation, so that we minimize biases arising from such procedure.

\begin{figure*}[t]  
  \vspace{0.0cm} \centerline{ \epsfysize=3.5truein
  \epsffile{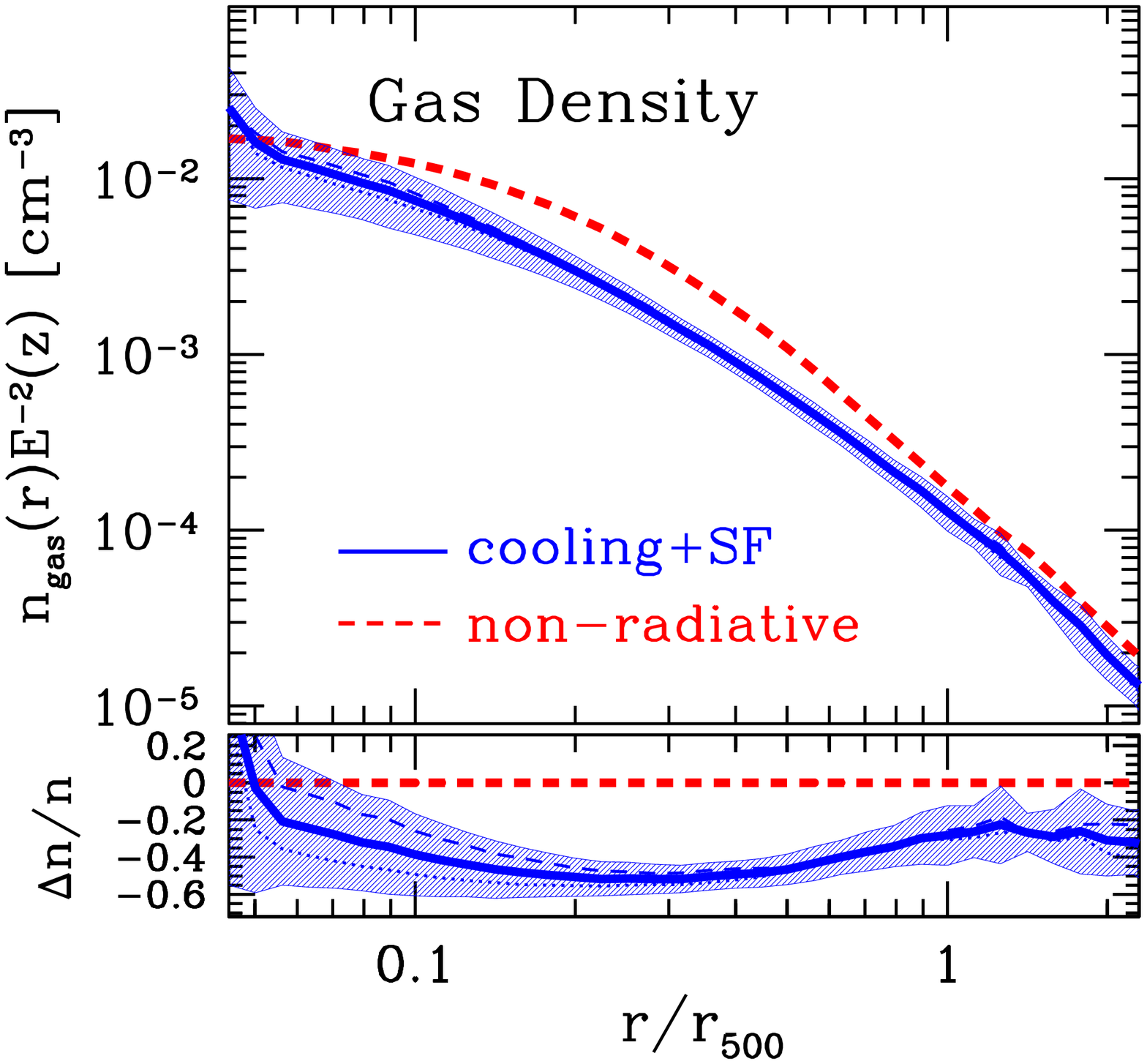} \hspace{-0.4cm} \epsfysize=3.5truein
  \epsffile{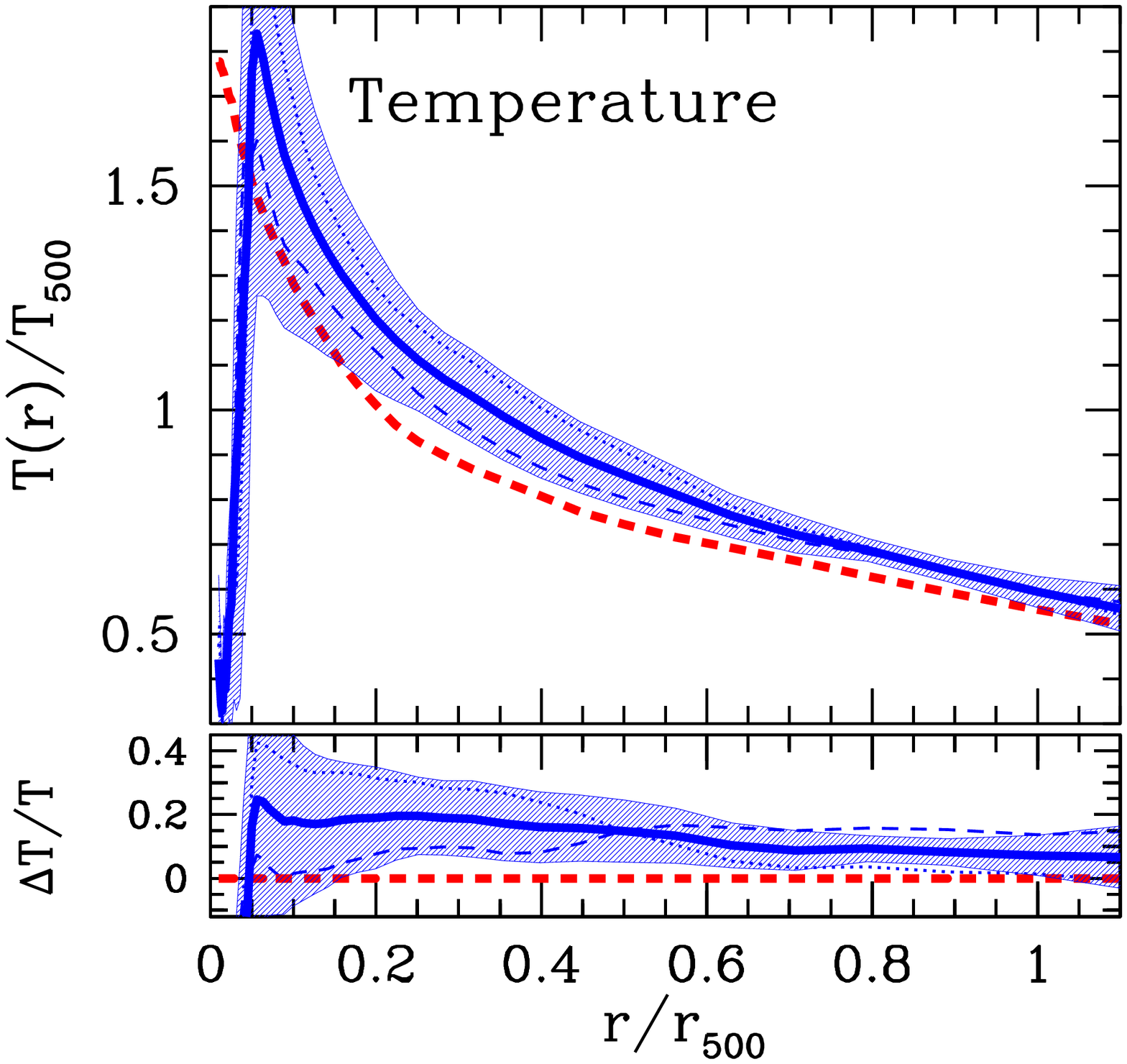} } \vspace{-0.8cm} \centerline{
  \epsfysize=3.5truein \epsffile{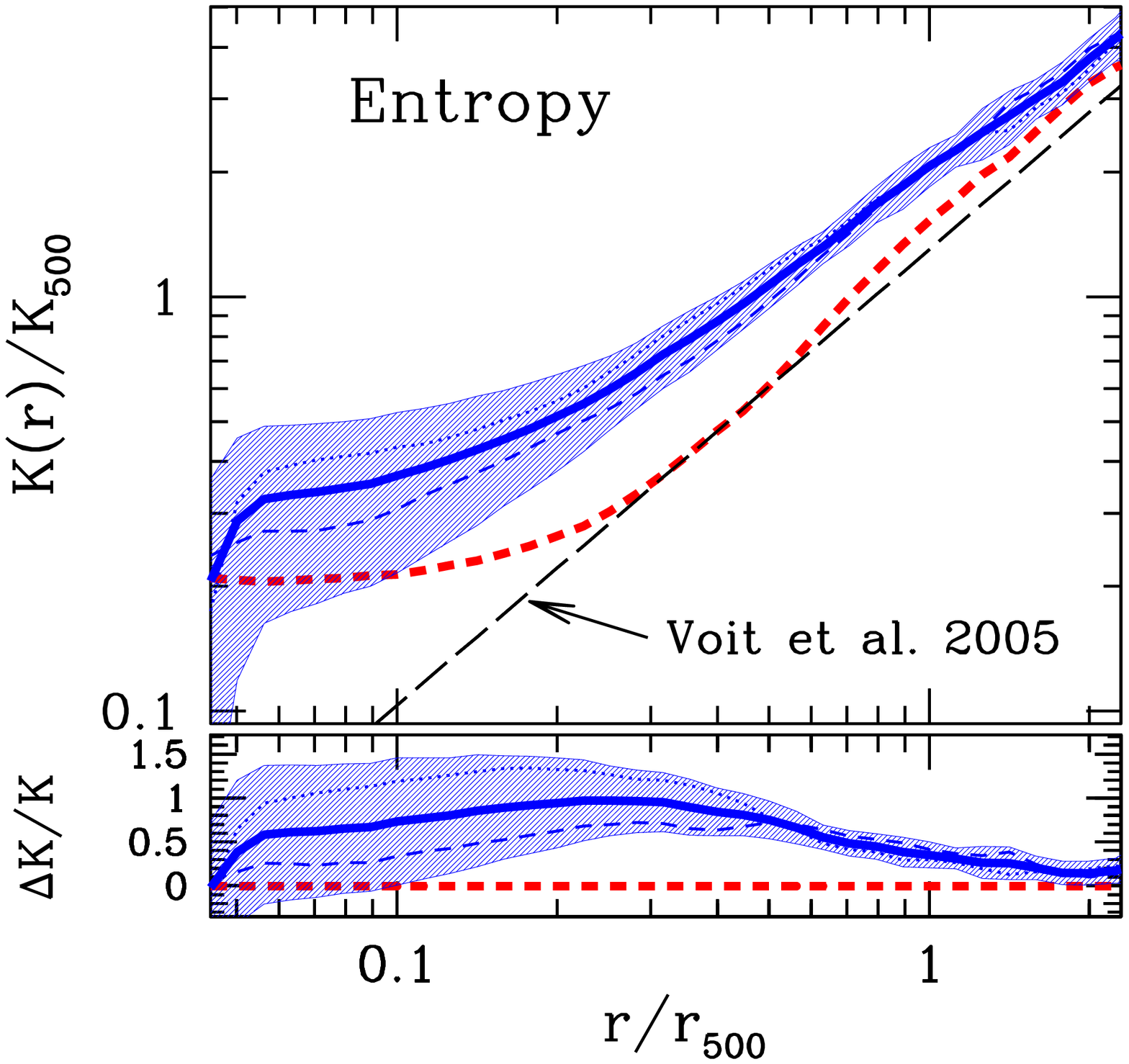} \hspace{-0.4cm}
  \epsfysize=3.5truein \epsffile{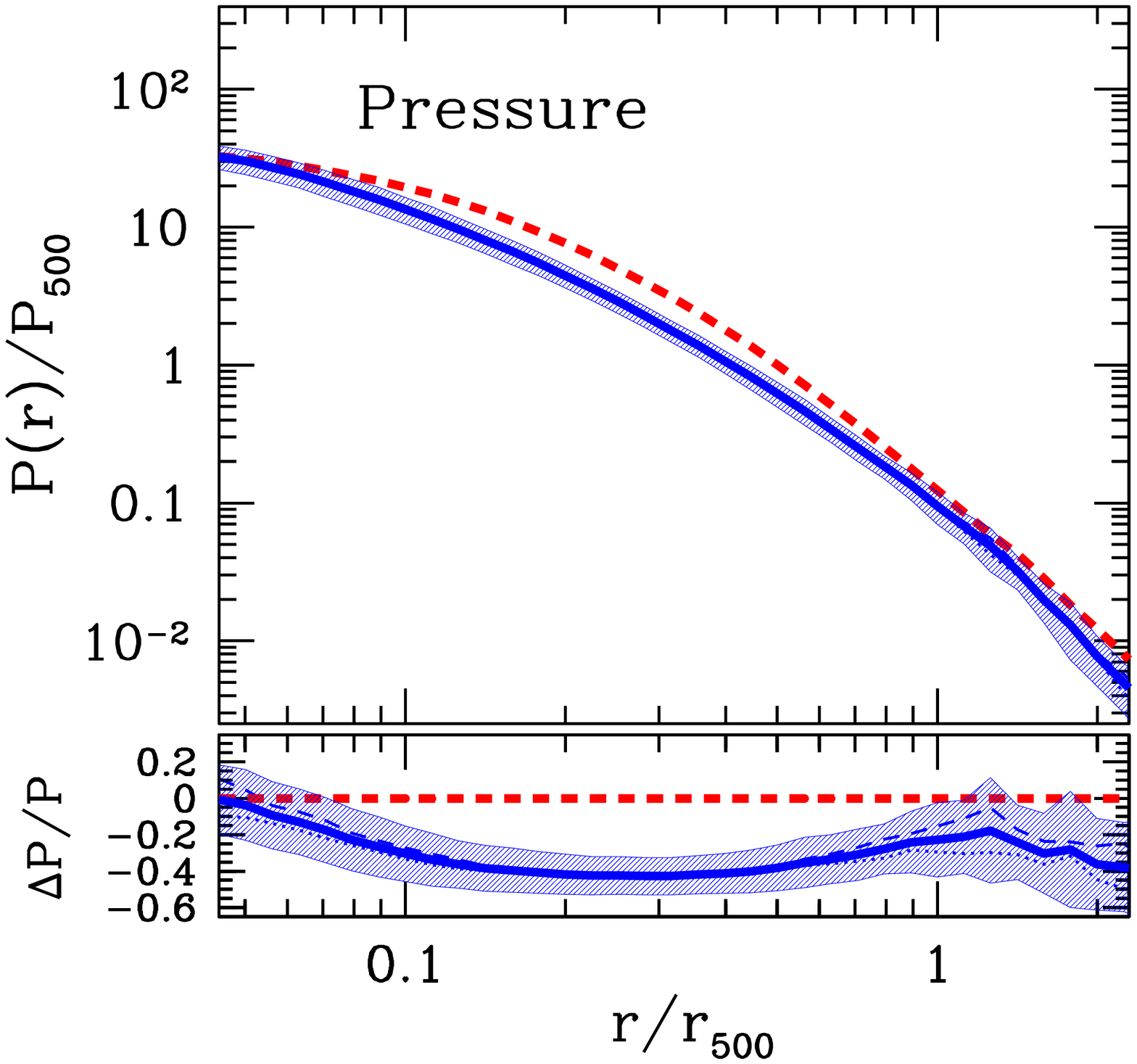} }
  \vspace{-0.4cm}
  \caption{ Radial profiles of the ICM in {\it relaxed} simulated
    clusters at $z=0$.  For each of the physical profiles {\it the
    upper panels} show the profiles, while the bottom panels show the
    corresponding fractional deviations of the profiles in the CSF
    simulation from the corresponding profiles in the non-radiative
    runs. The figure shows gas density ({\it top-left}), temperature
    ({\it top-right}), entropy ({\it bottom-left}), and pressure ({\it
    bottom-right}) profiles. Thick solid and dashed lines show the
    average profiles of the relaxed clusters in the CSF and
    non-radiative runs, respectively. The shaded band indicates the
    rms scatter around the mean profile for the CSF run.  In addition,
    the dashed and dotted lines indicate the average profiles of
    systems with $T_X>2.5$ and $<2.5$~keV, respectively, in the CSF
    simulations. Note that the entropy profiles of the non-radiative
    runs outside $0.3r_{500}$ are well-described by a power-law radial
    profile $K\propto r^{1.2}$, indicated by the dashed line in the
    bottom-left panel. }
\label{fig:pro_sim}
\end{figure*}

\begin{figure*}[t]  
  \vspace{0.0cm}
  \centerline{
    \epsfysize=3.5truein \epsffile{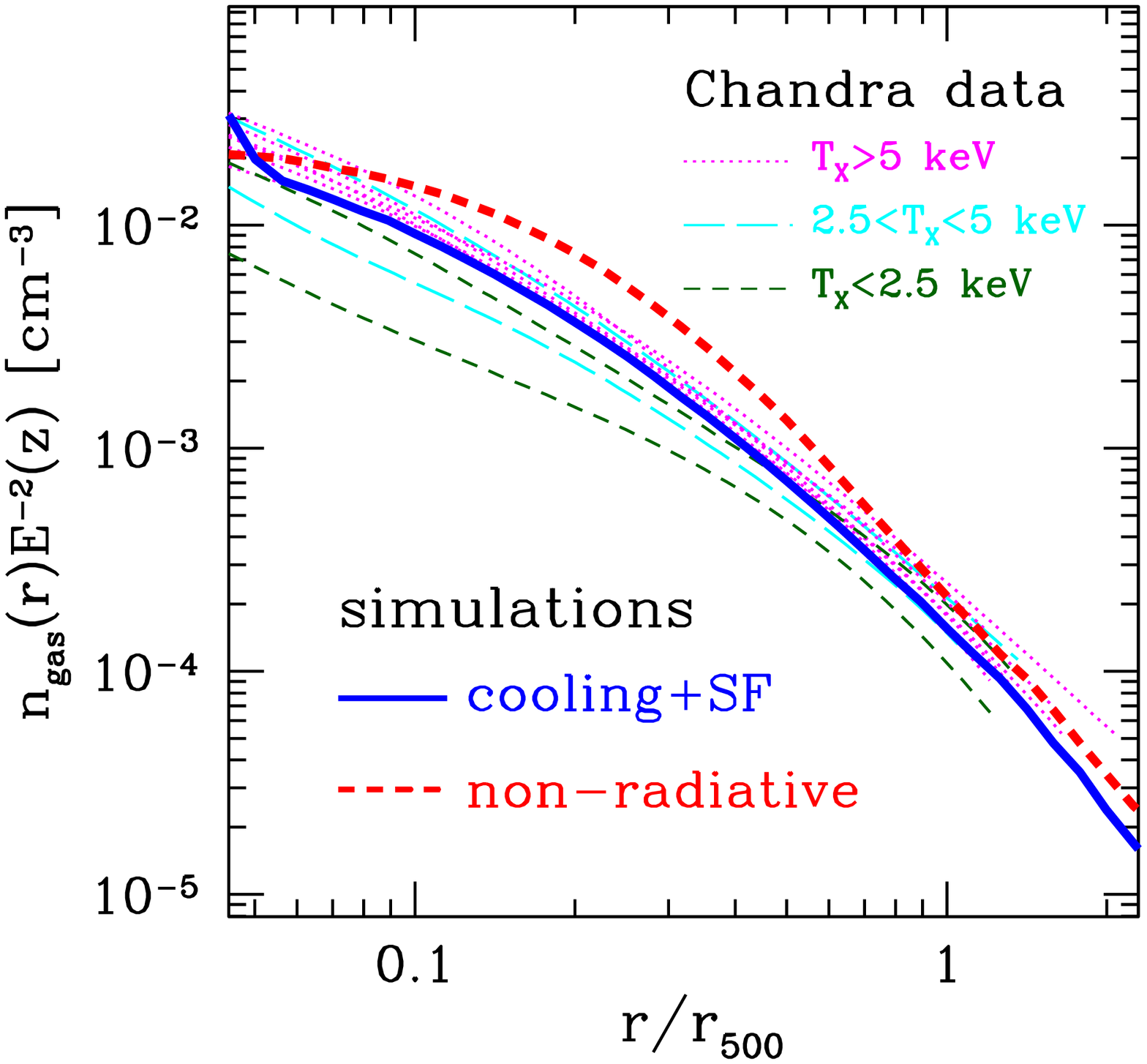} \hspace{-0.4cm}
    \epsfysize=3.5truein \epsffile{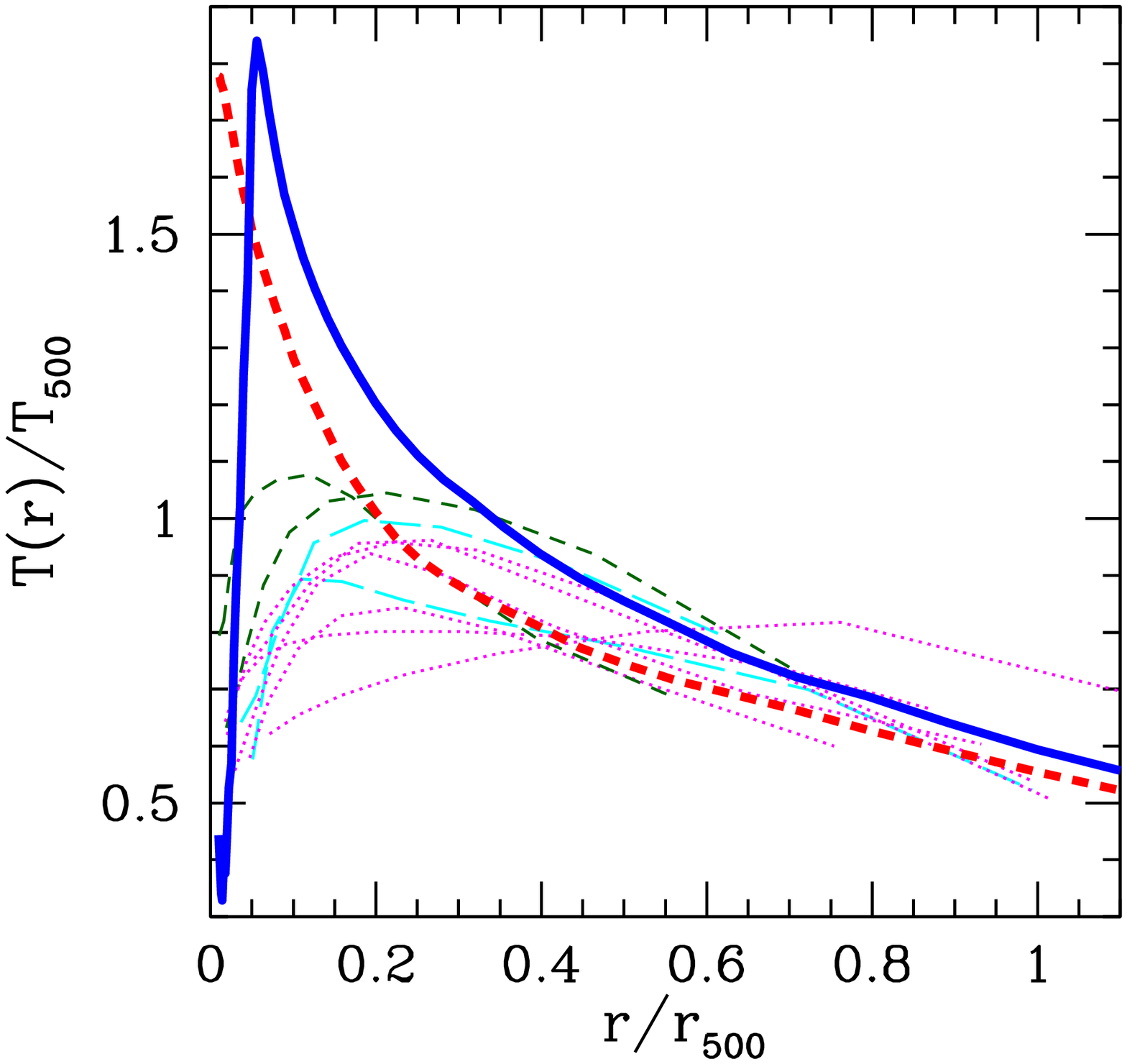}
  } \vspace{-0.8cm}
  \centerline{ 
    \epsfysize=3.5truein \epsffile{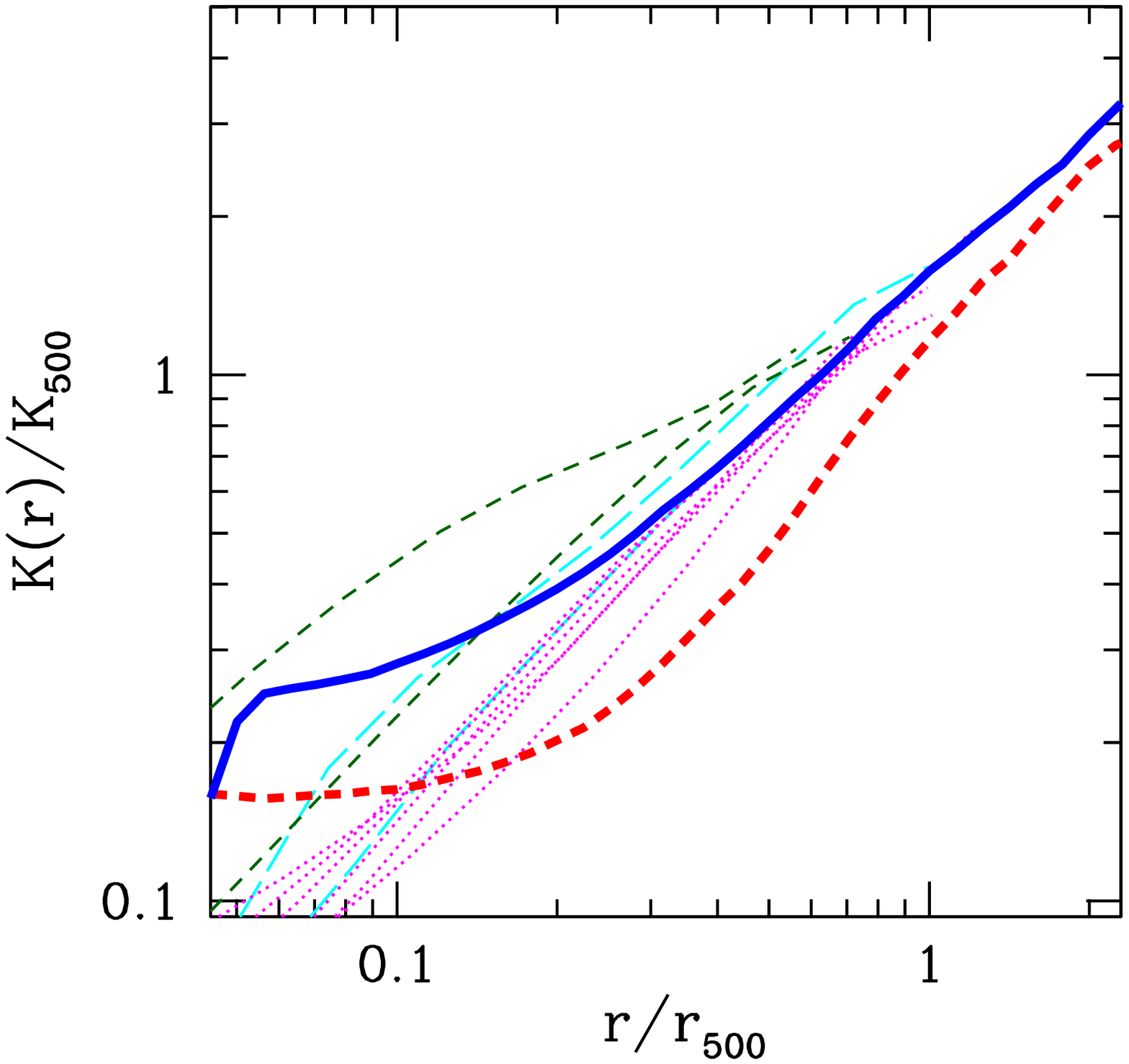} \hspace{-0.4cm} 
    \epsfysize=3.5truein \epsffile{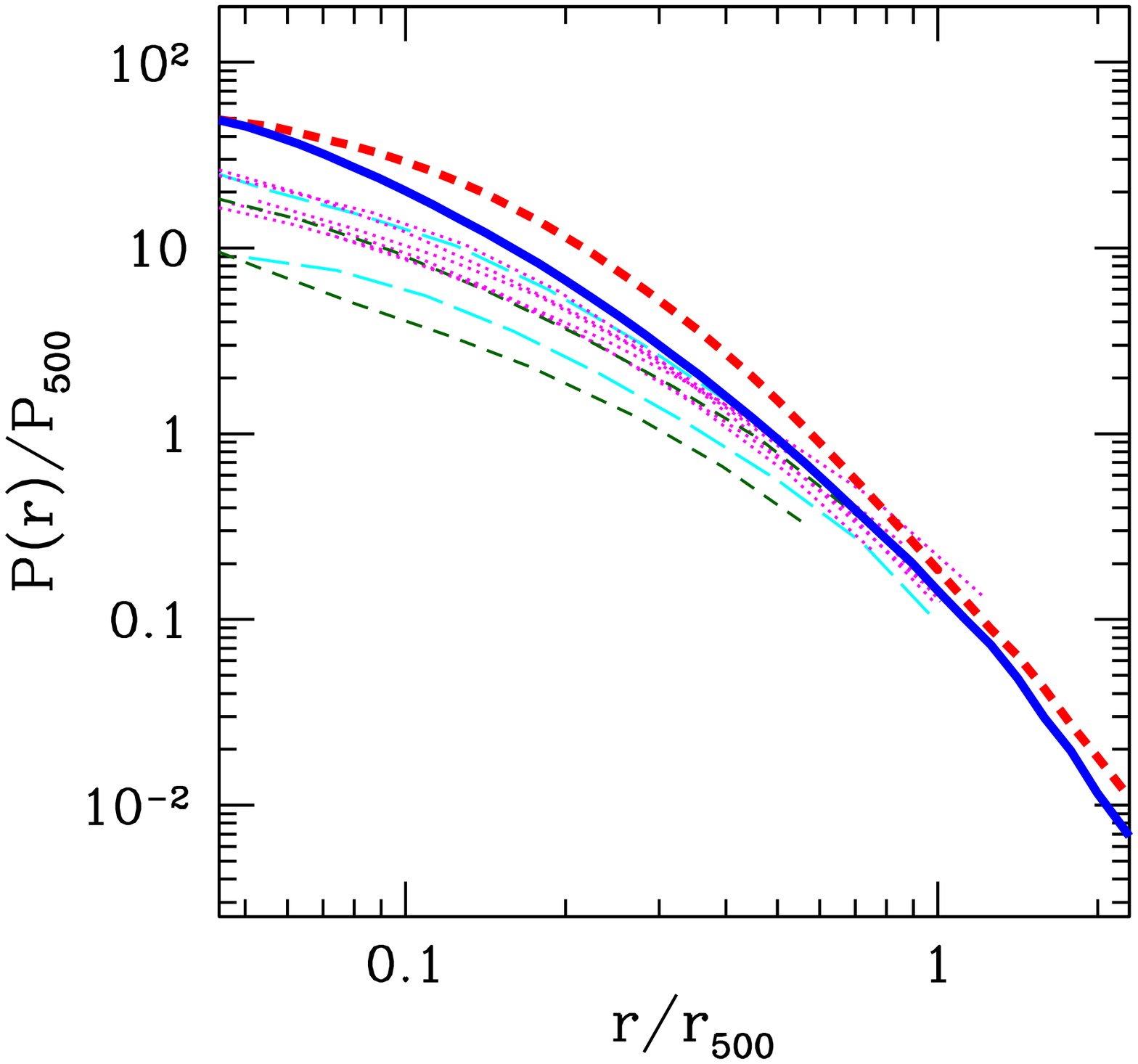} 
  }
  \vspace{-0.4cm}  
  \caption{Comparison of the ICM profiles in relaxed clusters at the
    present day ($z\approx 0$) in cosmological cluster simulations and
    the \emph{Chandra} sample of \citet{vikhlinin_etal06}.  The panels
    show the gas density ({\it top-left}), temperature ({\it
    top-right}), entropy ({\it bottom-left}), and pressure ({\it
    bottom-right}).  Thick solid and dashed lines show the mean
    profiles in the CSF and non-radiative simulations, respectively,
    while the observed profiles are shown by the thin dotted,
    long-dashed and short-dashed lines for the systems with $T_{\rm
    X}>5$~keV, $2.5<T_{\rm X}<5$~keV, and $T_{\rm X}<2.5$~keV,
    respectively. Note that at $r\gtrsim 0.1r_{500}$ the profiles of
    the CSF simulations provide a better match to the observed
    profiles than the profiles in the non-radiative runs. }
\label{fig:pro_comp}
\end{figure*}

\section{Effects of Galaxy Formation on the ICM Profiles}
\label{sec:profiles}
\label{sec:profiles_sim}

In this section we investigate the effects of galaxy formation on the
ICM properties by comparing simulations performed with and without the
processes associated with the galaxy formation: gas cooling, star
formation, stellar feedback, and metal enrichment.  Similar
comparisons have been done for a subset of 11 clusters in
\citet{kravtsov_etal05} and \citet{nagai06}.  Here, we use the
extended sample of 16 clusters and analyze the sub-sample of six
relaxed clusters that are identified as ``relaxed'' in all three
orthogonal projections, indicated as 111 in the last column of the
Table~\ref{tab:sim}.  In \S~\ref{sec:obs_comp}, we compare the results
of simulations to the \emph{Chandra} X-ray observations of nearby
relaxed clusters. 

Figure~\ref{fig:pro_sim} shows the average radial profiles of the ICM
in relaxed clusters at $z=0$ in the CSF and non-radiative runs.
Clockwise from the top-left panel, we show the gas density,
temperature, pressure, and entropy profiles. The mean profiles are
obtained by first normalizing the ICM profiles of each cluster at
$r_{500}$ and then averaging over a sample of relaxed clusters. The
shaded bands show $1\sigma$ rms scatter around the mean profile of the
CSF runs, and the mean and scatter of the profiles are computed for a
logarithm of each thermodynamic quantity.  We also examine systems
with $\Tx>$2.5~keV and $<$2.5~keV separately to study the mass
dependence of the effects (and also the effects of cooling in the
bremsstrahlung- and line emission-dominated regimes). In the bottom
panel of each figure, we also show the fractional change of the ICM
profiles in the CSF runs relative to the non-radiative runs.

The temperature, entropy (defined as $K\equiv k_B T/n_e^{2/3}$), and
pressure profiles are normalized to the values computed for the given
cluster mass using a simple self-similar model \citep{kaiser86,voit05}:
\begin{eqnarray} 
T_{500} & = & 11.05\,{\rm keV}\; \left( \frac{M_{500}}{10^{15}\,h^{-1}M_{\odot}}
  \right)^{2/3} E(z)^{2/3} \label{eq:t500} \\ 
K_{500} & = & 1963\,{\rm keV \; cm^{-2}}\; \left( \frac{M_{500}}{10^{15}\,h^{-1}M_{\odot}}
  \right)^{2/3} E(z)^{-2/3} \label{eq:k500} \\ 
P_{500} & = & 1.45\times 10^{-11} \,{\rm erg \; cm^{-3}}\; \left( \frac{M_{500}}{10^{15}\,h^{-1}M_{\odot}}
  \right)^{2/3} E(z)^{8/3} \label{eq:p500} 
\end{eqnarray}
where \M500\ is a total cluster mass enclosed within \r500, $E^2(z) =
\Omega_M(1+z)^{3}+\Omega_{\Lambda}$ for a flat universe with a
cosmological constant assumed in our simulations. Numerical
coefficients in equations~(\ref{eq:k500}) and (\ref{eq:p500}) follow
from the definitions $K_{500}\equiv k_B T_{500}/n_{e,500}^{2/3}$ and
$P_{500}\equiv n_{g,500}\,k_B\,T_{500}$, where $n_{e,500} =
(\mu/\mu_e)\,n_{g,500} = 500\,f_b\,\rhocrit/(\mu_e m_p)$, $\rhocrit$
is the critical density of the universe, $f_b\equiv \Omega_M/\Omega_b$
is the mean baryon fraction in the Universe, $\mu$ is the mean
molecular weight, and $\mu_e$ is the mean molecular weight per free
electrons.  Note that we use $\mu=0.59$ and $\mu_e=1.14$ throughout
this work.

Figure~\ref{fig:pro_sim} shows that including the gas cooling and star
formation significantly modifies the ICM profiles throughout the
cluster volume.  The effect is larger in the inner region and for the
systems with lower $T_X$ (or the cluster mass).  Compared to the
non-radiative runs, the gas density in the CSF runs is reduced by
$\approx 50$ and $25\%$ at $0.3$ and $1.0\,r_{500}$, because a
fraction of gas is converted into stars. At small radii, we observe a
trend with cluster mass; for example, the suppression of the gas
density in the CSF runs at $r=0.1\,r_{\rm 500}$ is $\sim 50\%$ and
30\% for systems with $\Tx<2.5$~keV and $>2.5$~keV,
respectively. However, at $r>0.3r_{500}$ (or $r>0.15r_{\rm vir}$), our
simulations show very little systematic trend with $T_X$, indicating
that the clusters become self-similar in the outskirts even when the
cooling and star formation are turned on.

The ICM temperature profiles decline monotonically from $0.05 r_{500}$
outwards.  The shape of the temperature profiles are similar between
the non-radiative and CSF runs, but there is a clear offset in their
normalization. The temperature in the CSF runs is systematically
higher by 10\%--20\% outside the core, indicating that the net effect
of gas cooling and star formation is to increase the ICM temperature. 
The effects of gas cooling and star formation on the ICM temperature
show a stronger dependence on cluster mass than gas density. 

The ICM entropy provides one of the most fundamental insights into
physical processes determining the thermodynamics of the ICM, because
it is expected to a be a conserved quantity, modified only by shock
waves and ``non-adiabatic'' processes we are interested in
\citep[e.g., see][and references
therein]{voit_etal02,voit_etal05}. Figure~\ref{fig:pro_sim} shows that
the entropy profiles in our non-radiative simulations scale
self-similarly and are well-described by a power law $K\propto
r^{1.2}$, at $r>0.3r_{500}$, in agreement with previous studies
\citep{voit_etal05}.  Note, however, that there is a systematic
discrepancy between the predictions of the Eulerian and SPH codes at
small radii \citep{frenk_etal99,ascasibar_etal03}.  However, since the
primary focus of this paper is on the ICM properties outside the
cluster core, we leave a detailed analysis of this entropy discrepancy
for future work.

Using the average $K(r)$ profile from the non-radiative simulations as
a baseline, we study the effects of gas cooling and star formation on
the ICM entropy.  Compared to the non-radiative runs, the ICM entropy
in the CSF runs is enhanced in the entire radial range of interest,
even at the virial radius $r_{\rm vir}\simeq 2\times r_{500}$.  This
is because cooling leads to the condensation of the lowest entropy
gas, which is replaced by the gas of higher entropy
\citep{bryan00,voit_bryan01}.  The effect strongly depends on radius
and is most pronounced in the inner regions with the largest effect of
$\sim 100\%$ near $r=0.3r_{500}$.  However, the entropy is enhanced by
$\sim 40\%$ even at $r=r_{500}$.  The figure also shows that the
magnitude of the effect in the inner regions depends on cluster mass,
but is approximately the same for all masses at $r \gtrsim 0.5
r_{500})$, indicating that cooling preserved self-similarity of the
cluster outskirts.

Finally, pressure profiles exhibit the most remarkable degree of
self-similarity and low-level of cluster-to-cluster scatter. Notice
that the average pressure profiles of low and high-$\Tx$ systems are
nearly identical.  This indicates that the self-similarity is best
preserved for the quantities directly proportional to the ICM pressure
or thermal energy, such as the integrated pressure $Y_{\rm SZ}\propto
\Mg T_{\rm mg}$ \citep{nagai06} and $Y_{X} \equiv \Mg T_{X}$
\citep{kravtsov_etal06}. Note, however, that inclusion of gas cooling
and star formation modifies the overall shape and normalization of the
pressure profiles and hence the $Y_{\rm SZ}$ and $Y_X$ parameters for
the clusters of a fixed mass. In our simulations, the ICM pressure is
suppressed by about 25\% and 40\% at $r_{500}$ and $r_{2500}$,
respectively. 

\begin{figure}[t]  
  \centerline{ \epsfysize=4.0truein \epsffile{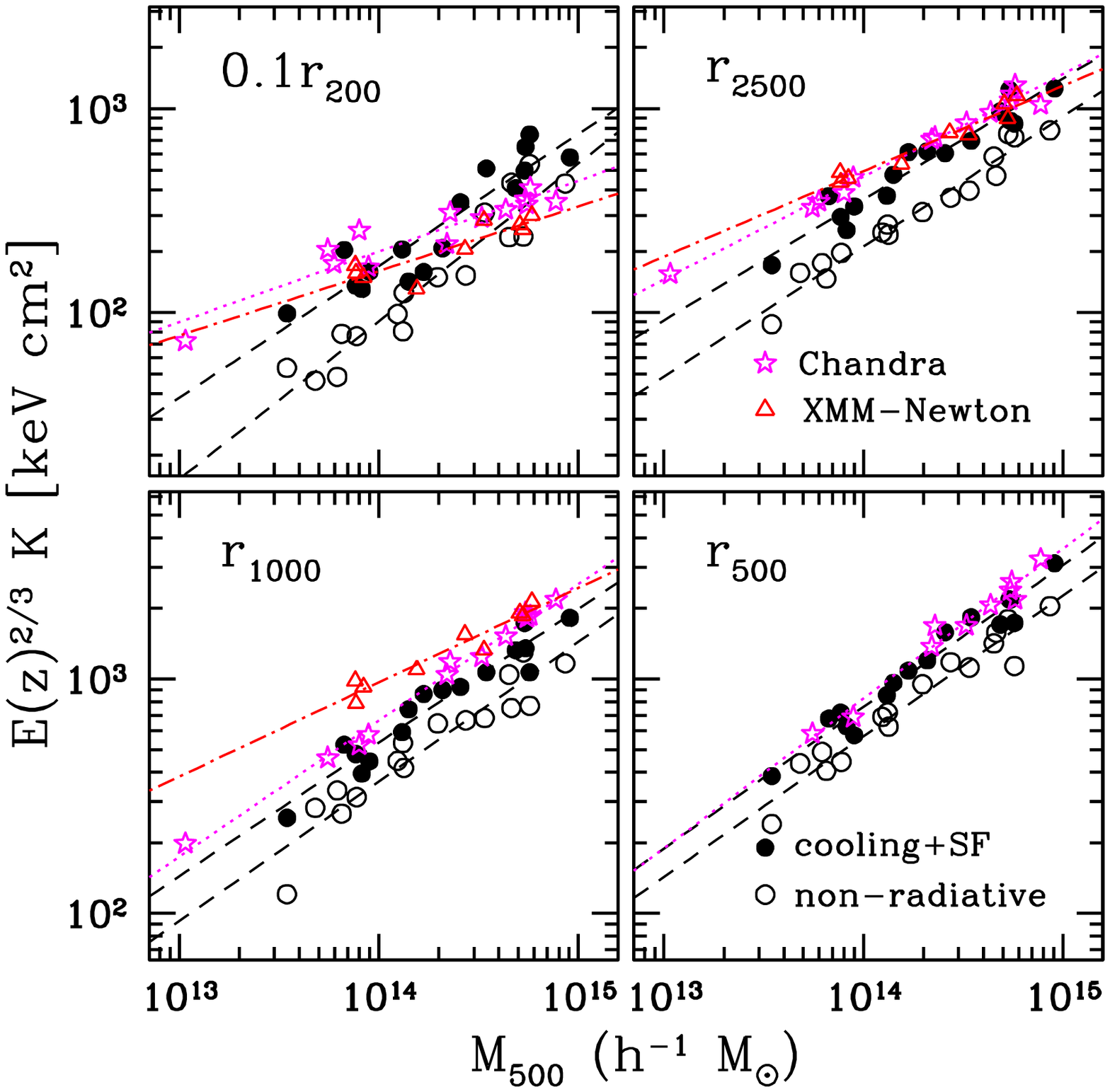} }
  \vspace{-0.7cm}
  \caption{ Correlation between the entropy $K\equiv k_B T/n_e^{2/3}$
    as a function of $\M500$. The entropy scaling relations are
    measured at $0.1 r_{200}$, $r_{2500}$, $r_{1000}$, and $r_{500}$.
    We compare the relations in the CSF and non-radiative simulations,
    indicated with filled and open circles, and the dashed and dotted
    lines indicate the best-fit power law relations to these sets of
    simulations, respectively.  Stars and triangles are observations
    by \emph{Chandra} \citep{vikhlinin_etal06} and \emph{XMM-Newton}
    \citep{pratt_etal06}.  The dashed lines indicate the best-fit
    power law relations to the CSF ({\it upper line}) and
    non-radiative ({\it lower line}) simulations, while the dotted and
    dot-dashed lines indicate the fits to the \emph{Chandra}, and
    \emph{XMM-Newton} measurements, respectively. }
\label{fig:K-M}
\end{figure}

\begin{figure}[t]  
  \centerline{ \epsfysize=4.0truein \epsffile{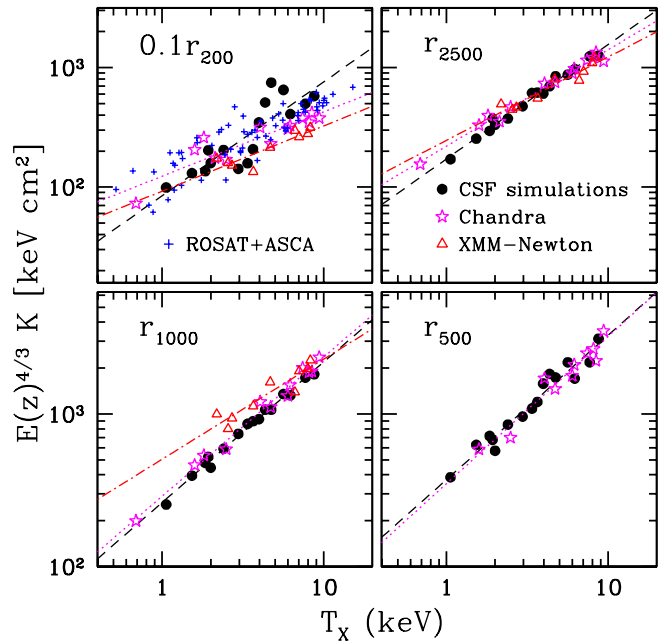} }
  \vspace{-0.7cm}
  \caption{ Correlation between the entropy $K\equiv k_B T/n_e^{2/3}$
    as a function of $\Tx$.  The entropy scaling relations are
    measured at $0.1 r_{200}$, $r_{2500}$, $r_{1000}$, and
    $r_{500}$. Solid circles indicate the CSF simulations, while
    stars, triangles, and crosses are \emph{Chandra}
    \citep{vikhlinin_etal06}, \emph{XMM-Newton} \citep{pratt_etal06},
    and \emph{ROSAT+ASCA} \citep{ponman_etal03} measurements.  The
    dashed, dotted, and dot-dashed lines indicate the best-fit power
    law relations to the CSF simulations, \emph{Chandra}, and
    \emph{XMM-Newton} measurements, respectively.}
\label{fig:K-Tx}
\end{figure}

\section{Comparisons with observations}
\label{sec:obs_comp}

In this section we present detailed comparisons of gas density,
temperature, entropy, and pressure profiles in the simulated clusters
and the \emph{Chandra} observations of low-$z$ relaxed clusters
\citep{vikhlinin_etal06}. We also compare the scaling relations
between $\Tx$, $\Mg$, the X-ray ``pressure'' ($\Yx\equiv \Mg\Tx$), and
cluster total mass.

\subsection{Profile comparison}
\label{sec:profiles_comp}

Figure~\ref{fig:pro_comp} compares the ICM profiles in simulations
with those observed.  For the simulated clusters, we plot the mean ICM
profiles averaged over the samples of relaxed clusters in both
non-radiative and CSF runs. They are compared to the \emph{Chandra} X-ray
measurements of 11 nearby relaxed clusters.  The observed clusters
with $\Tx>5$~keV, $2.5<\Tx<5$~keV, and $\Tx>5$~keV are indicated with
thin dotted, long-dashed, and short-dashed lines. 

The comparisons show that the ICM profiles in the CSF simulations
agree quite well with observations outside the cores of clusters
($r\gtrsim r_{2500}$), while the non-radiative simulations predict
overall shape and normalization of the ICM profiles inconsistent with
observations at all radii. The observations indicate then that a
significant amount of cooling and condensation of gas out of hot ICM
phase has occurred in real clusters. 

The ICM profiles in the inner regions, on the other hand, are not
reproduced well in any of our simulations. The only exception is the
$n_{\rm gas}(r)$ profiles for $\Tx>5$~keV clusters, where the CSF runs
produce results consistent with observation down to $r\simeq
0.06\,r_{500}$. However, even for these clusters the discrepancies
between simulations and observations are evident in the $T(r)$ and
$P(r)$ profiles at $r<0.3r_{500}$. The shape of the entropy profiles
is different in the inner region of the simulated and observed
clusters. For example, the entropy profiles of the observed clusters
monotonically decrease with decreasing radius, while simulated entropy
profiles flatten substantially at $r\lesssim 0.2r_{500}$.

Note that the ICM profiles of the $\Tx<2.5$~keV \emph{Chandra}
clusters are systematically offset from the high-$T_X$ clusters in the
inner regions and exhibit more pronounced cluster-to-cluster
variations. The lowest-$\Tx$ system, MKW4 ($\Tx=1.64$~keV), shows the
most striking deviations from self-similarity.  Our simulated clusters
show similar trends, but the sample is too small to quantify the
trends with $\Tx$ and the scatter. Note also that, if we use the
estimated $r_{500}$ of simulated clusters to compare with data, the
simulation curves in Figure~\ref{fig:pro_comp} could slide to the
left, bringing the characteristic values plotted in
Figures~\ref{fig:K-M} and ~\ref{fig:K-Tx} down slightly ($\lesssim
8\%$; see \S~\ref{sec:scaling} for more discussions).

\begin{deluxetable*}{lccccc}
\tablecaption{Best-fit parameters for the entropy-mass relation,
    $E(z)^{2/3} K=C(\M500/2\times 10^{14} h^{-1} M_{\odot})^{\alpha}$ at
    $z=0$.\label{tab:entropy1}}
\tablehead{
\nls
\multicolumn{1}{c}{radius}&
\multicolumn{1}{c}{quantity}&
\multicolumn{1}{c}{cooling+SF} &
\multicolumn{1}{c}{non-radiative} &
\multicolumn{1}{c}{\emph{Chandra}} &
\multicolumn{1}{c}{\emph{XMM-Newton}} 
\nls
}
\startdata
\nls
$0.1 r_{200}$  & $\log_{10}C$  & $2.423\pm0.030$  & $2.189\pm0.025$  & $2.391\pm0.020$  & $2.286\pm0.033$ \\
               & $\alpha$      & $0.647\pm0.077$  & $0.774\pm0.062$  & $0.335\pm0.046$  & $0.304\pm0.080$ \\
\nlss
$r_{2500}$     & $\log_{10}C$  & $2.734\pm0.017$  & $2.516\pm0.012$  & $2.805\pm0.011$  & $2.807\pm0.014$ \\
               & $\alpha$      & $0.593\pm0.046$  & $0.638\pm0.031$  & $0.492\pm0.038$  & $0.407\pm0.044$ \\
\nlss
$r_{1000}$     & $\log_{10}C$  & $2.890\pm0.016$  & $2.741\pm0.017$  & $2.984\pm0.011$  & $3.094\pm0.013$ \\
               & $\alpha$      & $0.570\pm0.043$  & $0.596\pm0.039$  & $0.569\pm0.018$  & $0.390\pm0.046$ \\
\nls
$r_{500}$      & $\log_{10}C$  & $3.063\pm0.013$  & $2.936\pm0.013$  & $3.100\pm0.013$  & \nodata \\
               & $\alpha$      & $0.605\pm0.036$  & $0.601\pm0.037$  & $0.598\pm0.028$  & \nodata \\
\enddata
\end{deluxetable*}

\begin{deluxetable}{lcccc}
\tablecaption{Best-fit parameters for the entropy-temperature relation
  $E(z)^{4/3} K=C(\Tx/{\rm 5~keV})^{\alpha}$ at $z=0$.\label{tab:entropy2}}
\tablehead{
\nls
\multicolumn{1}{c}{radius}&
\multicolumn{1}{c}{quantity}&
\multicolumn{1}{c}{cooling+SF} &
\multicolumn{1}{c}{\emph{Chandra}} &
\multicolumn{1}{c}{\emph{XMM-Newton}} 
\nls
}
\startdata
\nls
$0.1 r_{200}$  & $\log_{10}C$  & $2.593\pm0.042$  & $2.437\pm0.019$ & $2.320\pm0.035$  \\  
               & $\alpha$      & $0.954\pm0.077$  & $0.502\pm0.084$ & $0.506\pm0.131$  \\  
\nlss				       						      
$r_{2500}$     & $\log_{10}C$  & $2.905\pm0.009$  & $2.876\pm0.013$ & $2.851\pm0.021$  \\  
               & $\alpha$      & $0.958\pm0.031$  & $0.769\pm0.054$ & $0.657\pm0.089$  \\  
\nlss	   		       						      
$r_{1000}$     & $\log_{10}C$  & $3.064\pm0.006$  & $3.067\pm0.015$ & $3.135\pm0.026$ \\	      
               & $\alpha$      & $0.922\pm0.026$  & $0.887\pm0.043$ & $0.616\pm0.106$ \\	      
\nls	   		       						      
$r_{500}$      & $\log_{10}C$  & $3.232\pm0.018$  & $3.186\pm0.018$ & \nodata \\	      
               & $\alpha$      & $0.946\pm0.060$  & $0.930\pm0.063$ & \nodata \\           
\enddata
\end{deluxetable}

\subsection{Entropy scaling relations}
\label{sec:entropy_comp}

The scaling of the entropy with mass or $\Tx$ of clusters provides one
of the most powerful diagnostics of the effects of galaxy formation on
the ICM and deviations from self-similarity
\citep{evrard_henry91,ponman_etal99,voit_bryan01,pratt_etal06}.  For
the self-similar cluster model, the entropy at a fixed overdensity
radius is expected to scale linearly with $\Tx$ and with mass as
$\propto M^{2/3}$ (cf.{} eq.~\ref{eq:t500}). In the discussion above
we showed that entropy profiles of both observed and simulated
clusters become approximately self-similar outside the cluster cores.
In this section we explicitly consider the scaling of entropy with
cluster mass and temperature at four different radii and compare the
results of numerical simulations with X-ray measurements obtained
using \emph{Chandra} \citep{vikhlinin_etal06}. We also compare the
\emph{Chandra} measurements with the \emph{XMM-Newton}
\citep{pratt_etal06} and \emph{ROSAT+ASCA} \citep{ponman_etal03}
results.

The entropy levels measured in the simulated clusters at $0.1
r_{200}$, $r_{2500}$, $r_{1000}$, and $r_{500}$ are shown as a
function of $M_{500}$ in Figure~\ref{fig:K-M} \citep[such a relation
was first studied observationally by][]{pratt_etal06}. The best-fit
power law approximations of these data, $E(z)^{2/3}
K=C(M_{500}/2\times 10^{14}h^{-1}M_{\odot})^{\alpha}$, are given in
Table~\ref{tab:entropy1} (the $E(z)$ term corrects for evolution in
the self-similar model, which needs to be applied to clusters observed
at $z\ne0$).  Clearly, in the simulated clusters, inclusion of gas
cooling and star formation increases the entropy level, and the
magnitude of the effect is larger in the inner region.  The changes in
the normalization for $M=2\times 10^{14} h^{-1}M_{\odot}$ clusters is
a factor of 1.71, 1.65, 1.41, and 1.34 at $r=0.1\,r_{200}$,
$r_{2500}$, $r_{1000}$, and $r_{500}$, respectively. The effects of
cooling and star formation on the ICM entropy are thus stronger at
small radii. The slopes are consistent with the prediction of the
self-similar model ($\alpha = 2/3$) within 1 $\sigma$ at all radii, in
both CSF and non-radiative simulations.  However, there are
indications that the slopes are slightly shallower than the
self-similar value, and the slope in the non-radiative run is somewhat
steeper than that in the CSF at $r<r_{500}$. A larger sample of
simulated clusters is needed to determine whether these differences
are real.

It is easier to compare the entropy levels in the simulated and
observed clusters via the $K-T_X$ correlation. The results for our CSF
runs are $r=0.1\,r_{200}$, $r_{2500}$, $r_{1000}$, and $r_{500}$ are
shown in Figure~\ref{fig:K-Tx}. Table~\ref{tab:entropy2} lists the
best-fit parameters of the power approximations, $E(z)^{4/3}
K=C(\Tx/5~{\rm keV})^{\alpha}$. As in the case of the $K-M$ relation,
the power-law slopes in the CSF runs are slightly shallower than, but
consistent with, the self-similar expectation ($\alpha=1$) at all
radii considered. Note also that the $K-\Tx$ relations exhibit
remarkably tight relations at $r\geq r_{2500}$ for all clusters.

We also show in Figure~\ref{fig:K-M} and~\ref{fig:K-Tx} the $K-M$ and
$K-T$ scalings derived from several sets of X-ray cluster
observables. It is most straightforward to compare our simulations
with the \emph{Chandra} results of \citet{vikhlinin_etal06} (shown by
stars in Figures~\ref{fig:K-M} and~\ref{fig:K-Tx}), because we
explicitly tested their data analysis procedures \citep{nagai_etal07}
and because the \emph{Chandra} results for many clusters extend to
$r_{500}$. First, we note the entropy normalizations in the
\emph{Chandra} clusters show a good overall agreement with the CSF
runs at all radii. The results for non-radiative runs are strongly
inconsistent with the data \citep[see
also][]{ponman_etal03,pratt_etal06}.

At small radii, $r_{2500}$ and $0.1r_{200}$, the observed clusters
show significantly shallower slopes than expected in the self-similar
model and seen in the simulations. For example, the slope of the $K-T$
relation at $r=0.1 r_{200}$ is $\alpha_{\rm CSF}=0.95\pm0.08$ for the
simulated clusters, and $\alpha_{\rm Chandra}=0.50\pm0.08$ for
\emph{Chandra} sample. These results are in line with the disagreement
between the $K(r)$ profiles of the simulated and observed clusters at
small radii, discussed in \S~\ref{sec:profiles_comp}.

The agreement, however, improves quickly as we go to larger radii. At
$r_{2500}$, the slopes of the $K-T$ relation are $\alpha_{\rm
CSF}=0.96\pm0.03$ for simulated clusters and $\alpha_{\rm
Chandra}=0.77\pm0.05$ for the \emph{Chandra} sample. At larger radii,
$r_{1000}$ and $r_{500}$, the \emph{Chandra}-observed relations are
fully consistent with the CSF simulations both in terms of slope and
normalization; the slopes are also very close to the self-similar
expectations, $\alpha_{K-M}=2/3$ and $\alpha_{K-T}=1$. These results
confirm the general conclusion of \S~\ref{sec:profiles_comp} that
although effects of cooling on the entropy normalization are
significant within radii as large as $r_{500}$, the scaling of the
thermodynamic properties of the ICM become close to the self-similar
expectation outside the inner cluster region.

Also shown in Figure~\ref{fig:K-M} and~\ref{fig:K-Tx} are the entropy
scaling relations derived from two more X-ray data sets, the
\emph{XMM-Newton} sample of \citet{pratt_etal06} [triangles], and the
\emph{ASCA}+\emph{ROSAT} sample of \citet{ponman_etal03} [crosses at
$r=0.1\,r_{200}$]. At small radii, where the X-ray measurements are
most straightforward, there is a good agreement between all observed
relation. In particular, the \emph{XMM-Newton} and \emph{Chandra}
relations for $r=0.1\,r_{200}$ and $r_{2500}$ are nearly identical. A
small offset of the \citet{ponman_etal03} data points can be explained
by the slightly different definitions of $r_{200}$ used in these
works\footnote{The quantity $r_{200}$ was defined in
\citet{ponman_etal03} through the \citet{evrard_etal96} scaling with
$T_X$; through the NFW fit to the data at smaller radii in
\citet{pratt_etal06}; and through hydrostatic estimates of $r_{500}$
for the \emph{Chandra} clusters.}.

At $r=r_{1000}$, the entropy normalizations for the most massive
clusters are in agreement for the \emph{Chandra} and \emph{XMM-Newton}
samples but there is some tension in the values of slope. The
\emph{XMM-Newton} results indicate nearly the same slopes at
$r=r_{1000}$ and smaller radii, all significantly flatter than the
self-similar prediction: $\alpha_{K-T}=0.51\pm0.13$, $0.66\pm0.09$,
and $0.62\pm0.11$ for $r=0.1\,r_{200}$, $r_{2500}$, and $r_{1000}$,
respectively. The \emph{Chandra} results clearly indicate a
significantly steeper slope at this radius, $0.89\pm0.04$; a steep
slope, $0.93\pm0.06$ is also observed at $r=r_{500}$. The statistical
significance of the difference between the \emph{Chandra} and
\emph{XMM-Newton} slopes is $3.4\sigma$ for the $K-M$ and $2.4$
$\sigma$ for the $K-T$ relations. The best-fit values of $\alpha$
indicate \emph{qualitatively} different cluster properties. While the
\emph{XMM-Newton} results suggest that the departures from
self-similar scalings are of the similar amplitude at all radii, the
\emph{Chandra} measurements clearly point in the direction that the
effect is confined to the very central regions.

A detailed comparison of the \emph{XMM-Newton} and \emph{Chandra} data
analyses is beyond the scope of this work. We only point out two
effects that may contribute to the difference in the entropy
scalings. First, the \emph{Chandra} temperature profiles show a
systematic decline at large radii (by a factor of $\sim 1.7$ at
$r=r_{500}$ relative to peak values near $\sim 0.2\,r_{500}$), the
\emph{XMM-Newton} temperature profiles in the \citet{pratt_etal06}
sample are much flatter. This systematic difference is discussed in
\citet{vikhlinin_etal05c}. Second, the gas densities in the
\emph{Chandra} analysis were derived using a model that allows for
steepening of the $\rho(r)$ profile at large radii. The
\emph{XMM-Newton} data were fit with the $\beta$-type models that do
not allow for such steepening \citep{pointecouteau_etal05}. This leads
to somewhat different gas density profiles at large radii \citep[see,
e.g., Appendix A2 in][]{vikhlinin_etal06}.

\begin{figure*}[t]
  \centerline{ \hspace{0.2cm} \epsfysize=2.5truein \epsffile{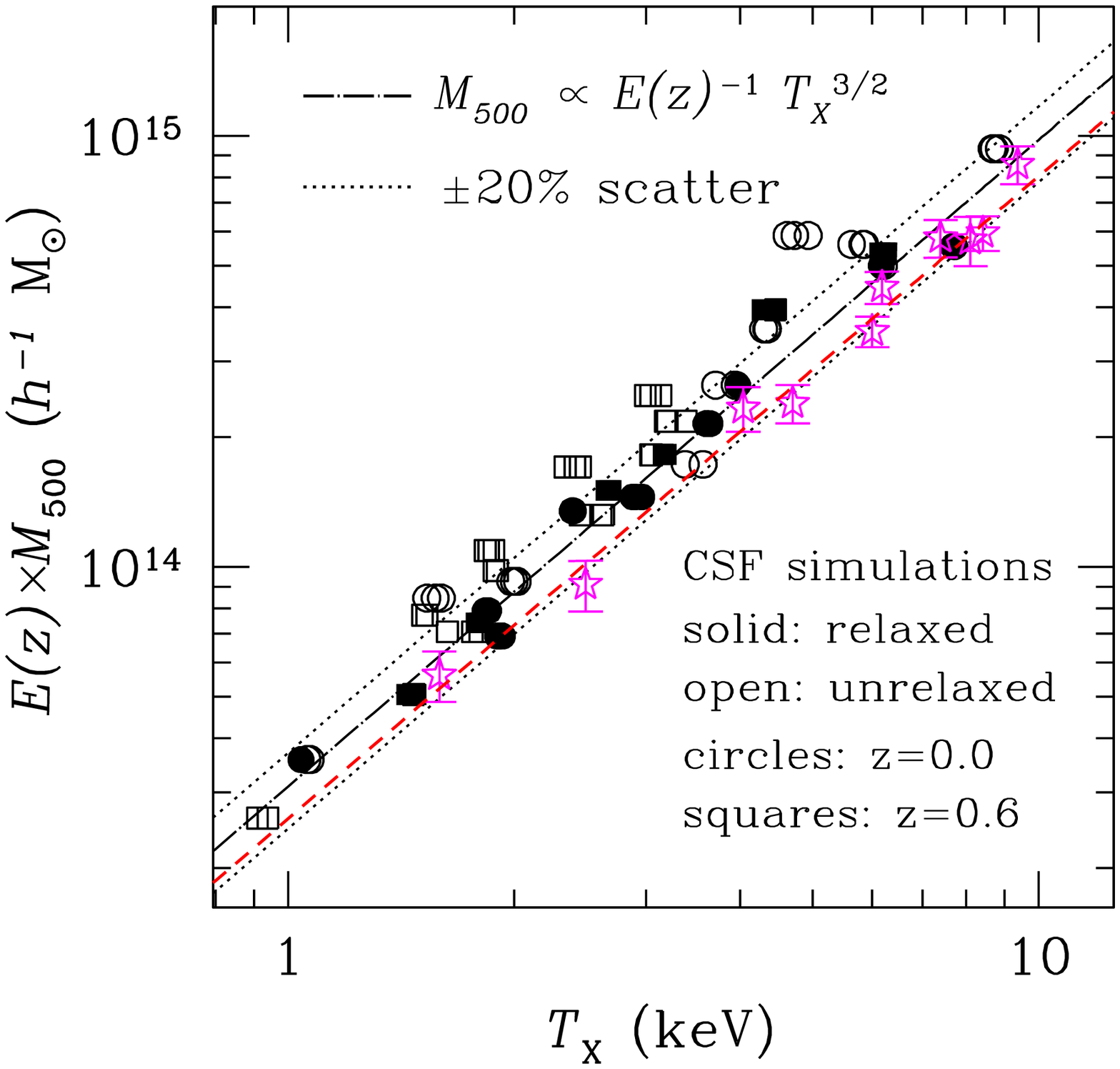}
    \hspace{-0.6cm} \epsfysize=2.5truein \epsffile{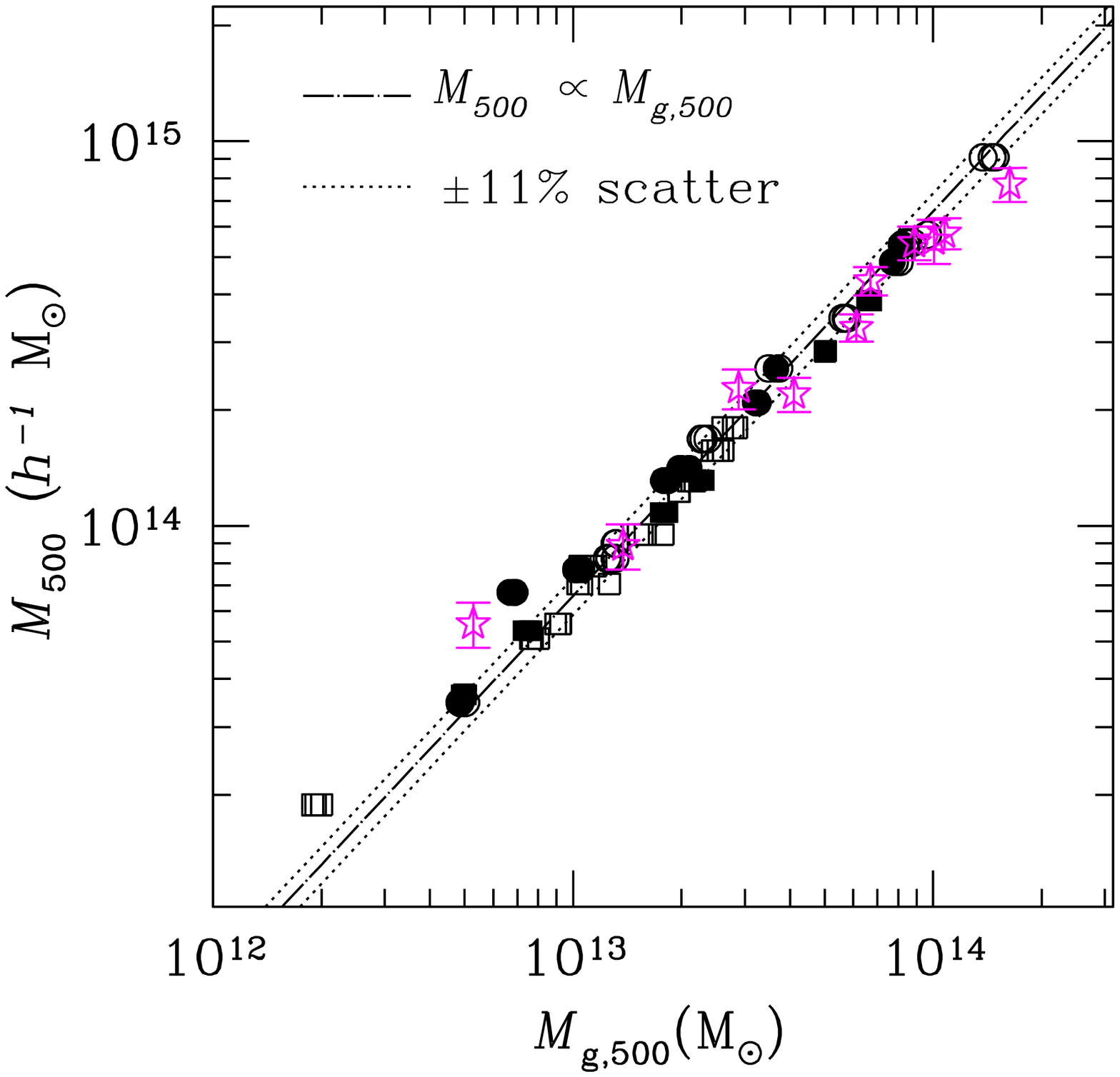} \hspace{-0.6cm}
    \epsfysize=2.5truein \epsffile{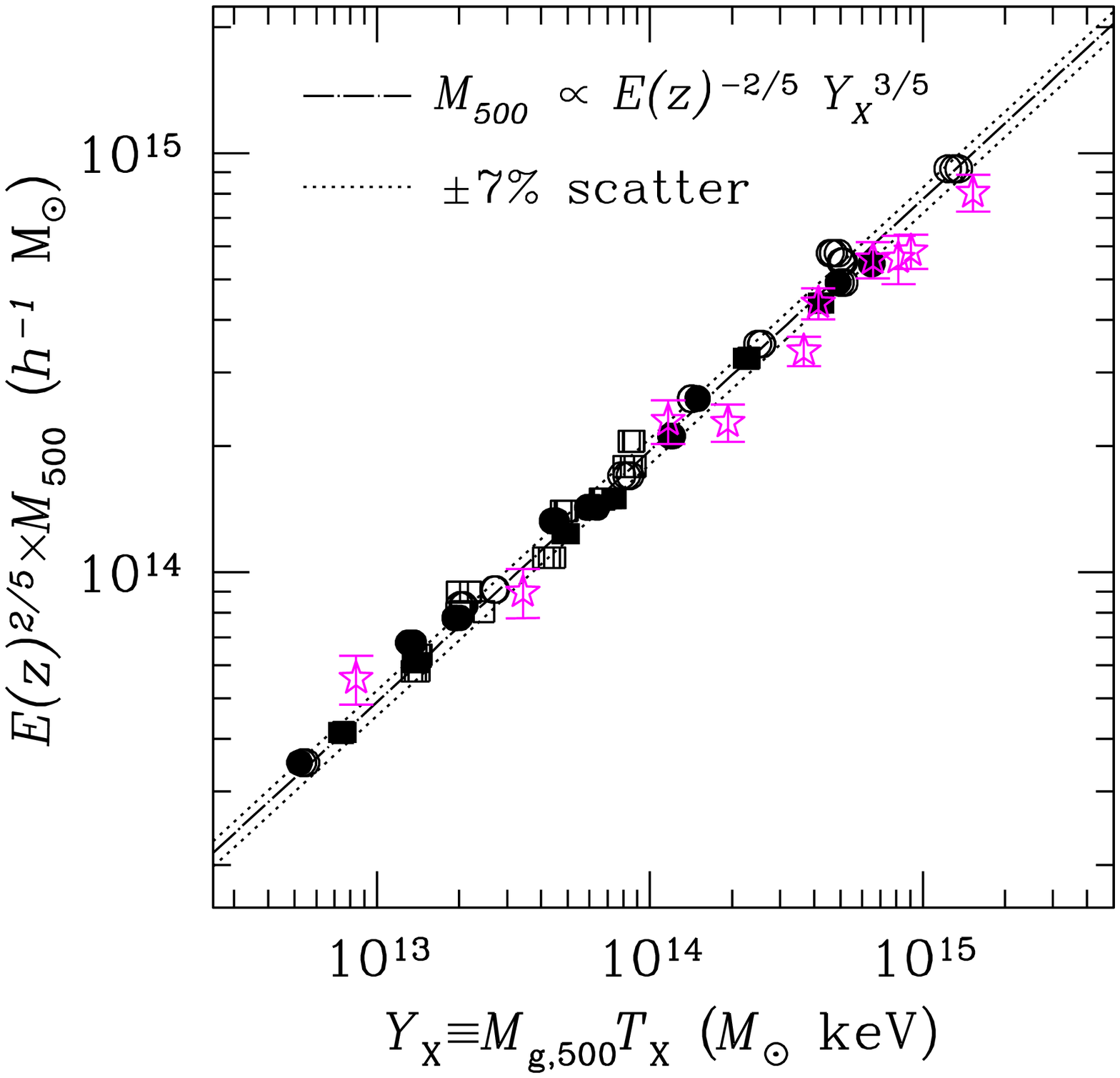}} \vspace{-0.6cm}
  \centerline{ \hspace{0.2cm} \epsfysize=2.5truein \epsffile{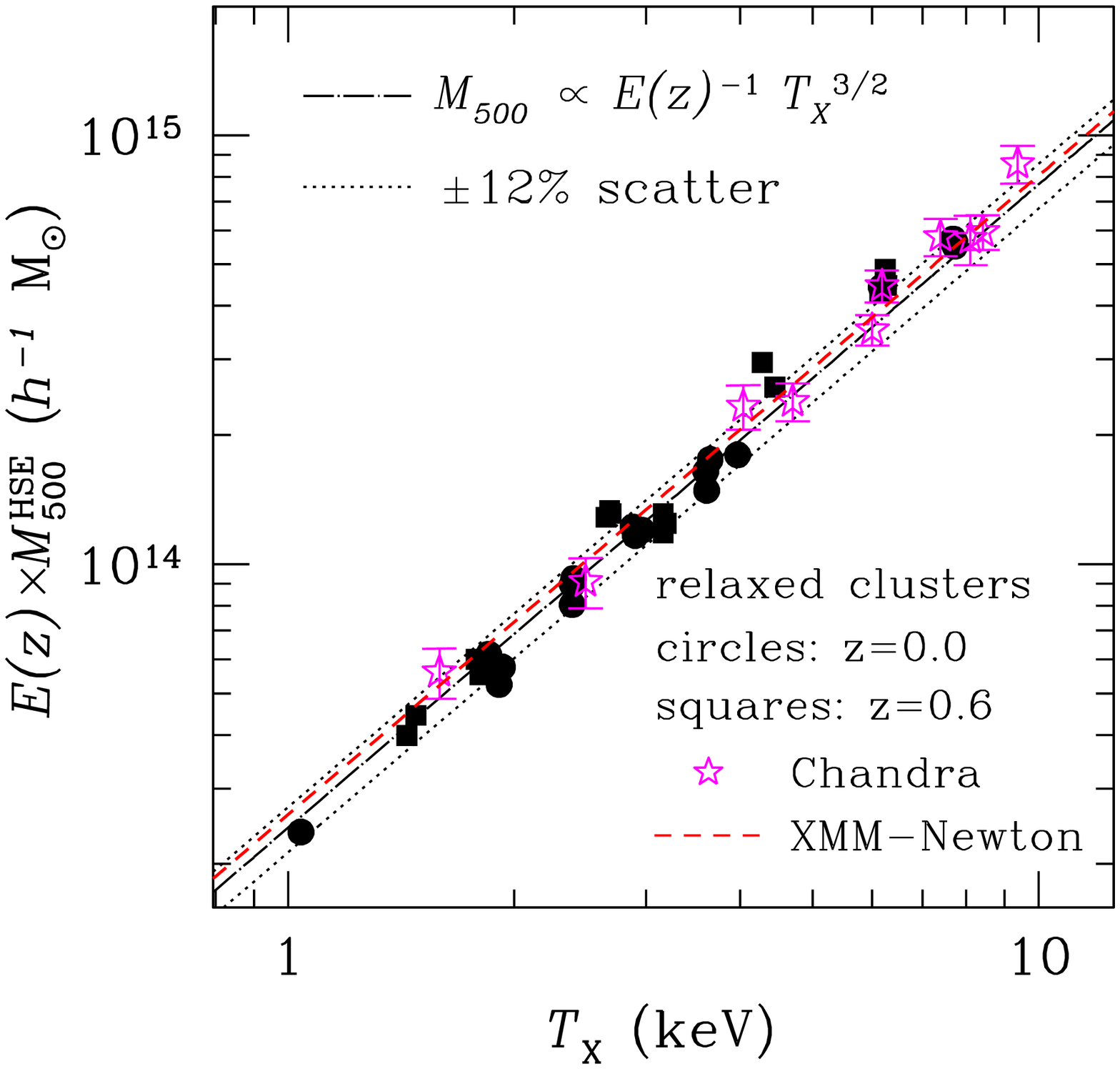}
    \hspace{-0.6cm} \epsfysize=2.5truein \epsffile{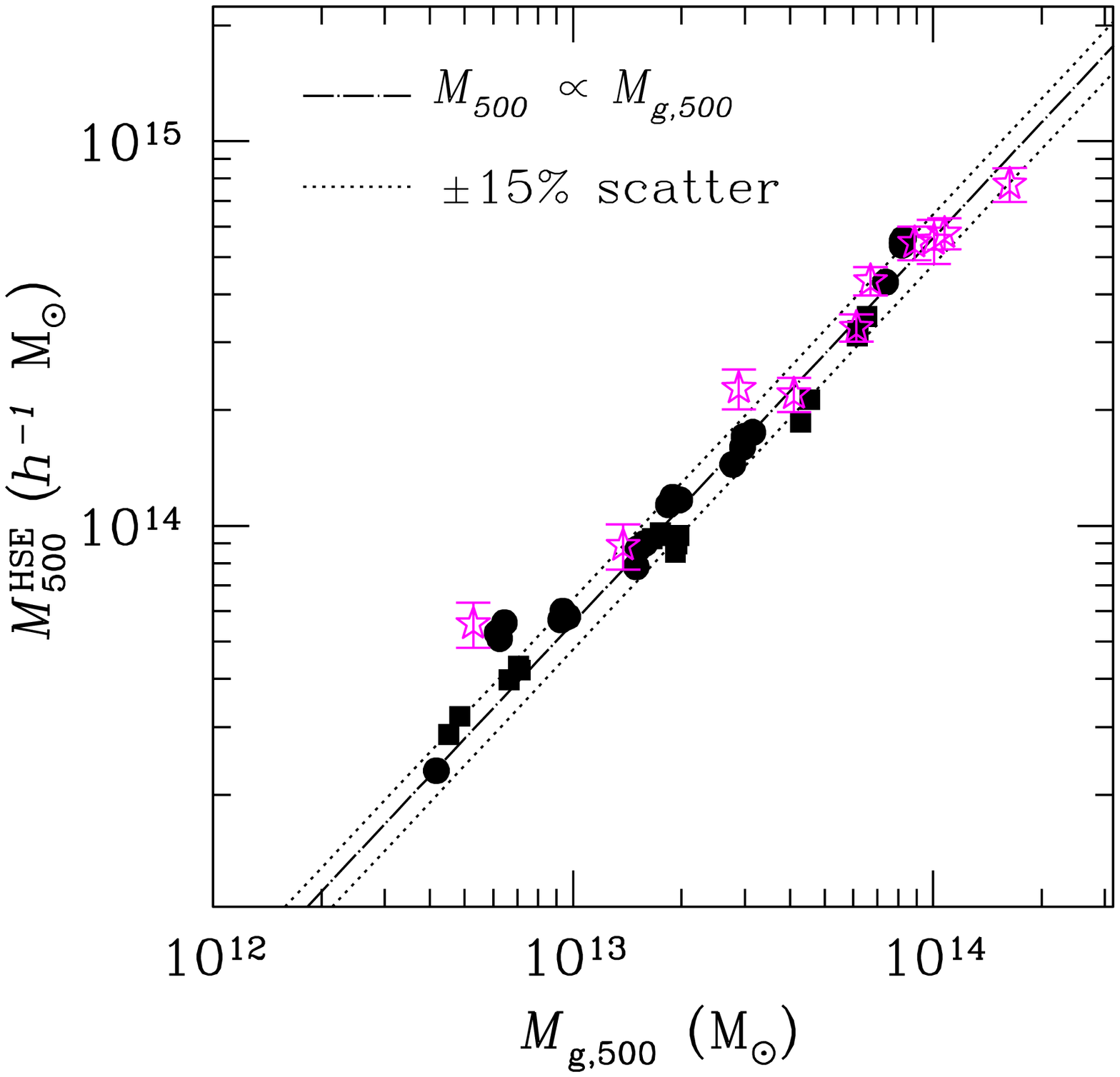} \hspace{-0.6cm}
    \epsfysize=2.5truein \epsffile{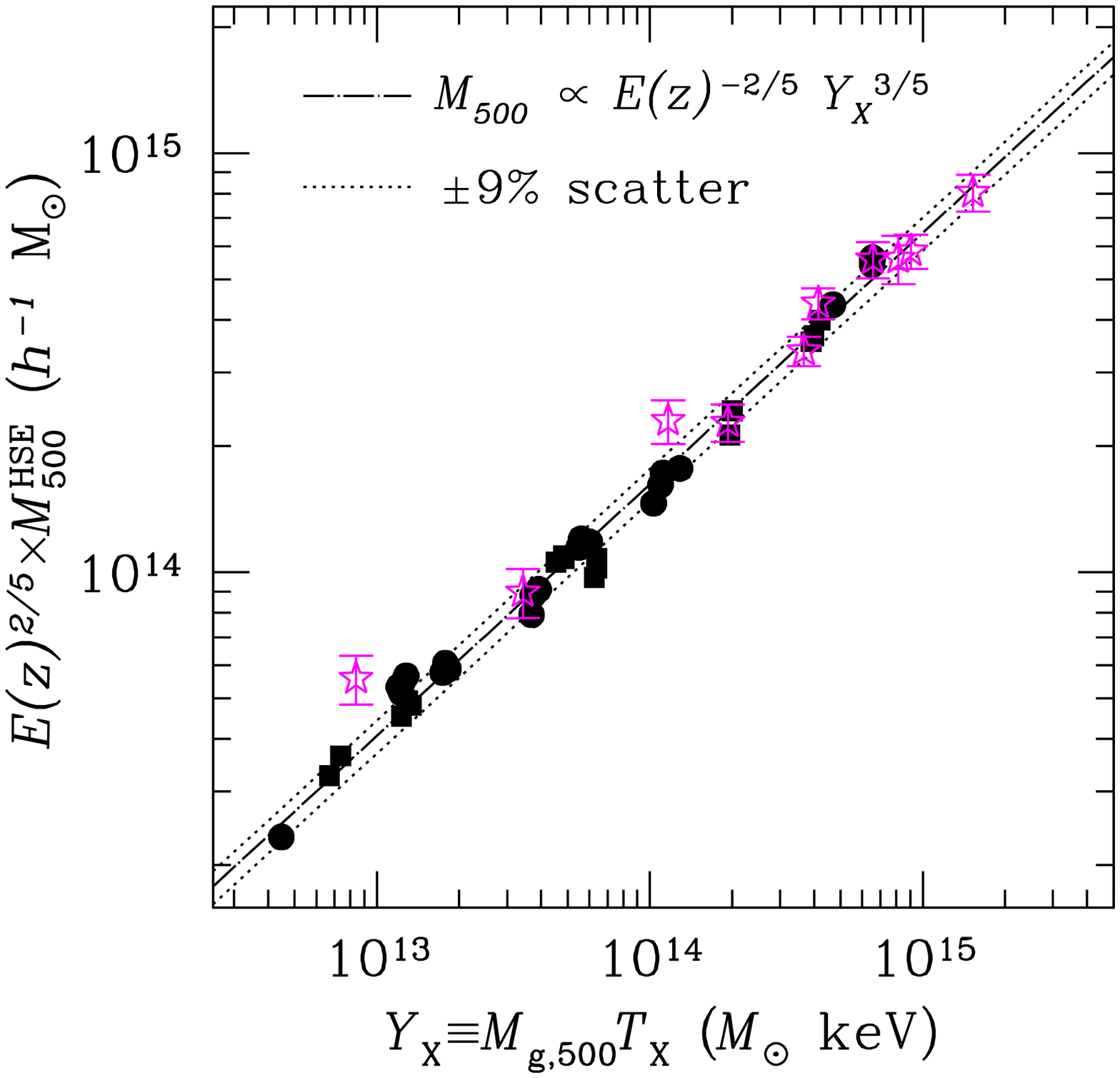}}
  \vspace{-0.2cm}
  \caption{ Comparisons of the X-ray observable-mass relations in
    simulations and observations. {\it From left to right:}
    Correlations between the total mass, $\M500$, and X-ray spectral
    temperature ($\Tx$), gas mass ($\Mg$), and X-ray pressure
    ($Y_X\equiv \Tx \Mg$). Relations are shown for the true 3D cluster
    mass $\M500\equiv M(<r_{500}^{\rm true})$ as measured in
    simulations ({\it upper panels}) and the hydrostatic mass $M^{\rm
    HSE}_{500}\equiv M^{\rm HSE}(<r_{500}^{\rm est})$ derived from
    mock \emph{Chandra} analysis ({\it lower panels}).  Separate
    symbols indicate relaxed and unrelaxed simulated clusters, and
    also $z=0$ and $0.6$ samples.  The figures include points
    corresponding to three projections of each cluster.  The
    dot-dashed lines are the power law relation with the self-similar
    slope fit for the sample of relaxed clusters.  The dotted lines
    indicate the rms scatter around the mean relation. The data points
    with error bars are \emph{Chandra} measurements of nearby relaxed
    clusters. The dashed line is the best-fit M-$T_X$ relation from
    the \emph{XMM-Newton} measurements.}
\label{fig:xscale}
\end{figure*}

\subsection{Relations between Total Mass and X-ray Observables}
\label{sec:scaling}

We present comparisons of the X-ray observable-mass relations of the
CSF simulations and \emph{Chandra} X-ray observations of nearby
relaxed clusters in Figure~\ref{fig:xscale}.  Following
\citet{kravtsov_etal06}, we consider three X-ray proxies for the
cluster mass: --- the spectral temperature ($\Tx$), the gas mass
($\Mg$), and the X-ray pressure ($Y_X\equiv \Tx \Mg$).  These X-ray
mass proxies are derived from mock \emph{Chandra} images of the
simulated clusters and analyzing them using a model and procedure
essentially identical to those used in real data analysis. Note that
the mean temperatures were estimated from a single-temperature to the
spectrum integrated in the radial range $[0.15-1]r_{500}$ (i.e.,
excluding emission from cluster core).

In the upper panels of Figure~\ref{fig:xscale}, we compare the scaling
relations of simulated clusters for the true cluster mass,
$M_{500}^{\rm true}(<\!r_{500}^{\rm true})$, measured in simulations
to the relations from the \emph{Chandra} X-ray cluster observations.
We also plot the best-fit $M-\Tx$ relation from the \emph{XMM-Newton}
measurements \citep{arnaud_etal05} for comparison.  Results of power
law fits to these relations for different subsets of the clusters are
presented in Table 2 of \citet{kravtsov_etal06}.\footnote{Note that
$\M500$ is in units of $h^{-1} M_{\odot}$ in this work, while it is
$M_{\odot}$ in \citet{kravtsov_etal06}.} The comparisons show that the
normalizations of the scaling relations involving true $M_{500}$ for
our simulated sample are systematically high by $\approx 10-20\%$
compared to the observed relations. We note that this level of
agreement is considerably better than agreement between simulations
and data as recently as several years ago
\citep[e.g.,][]{pierpaoli_etal01}.

The remaining bias could arise from the assumption of the hydrostatic
equilibrium, which is a key assumption that enables measurements of
gravitationally bound mass of clusters from the X-ray and
Sunyaev-Zel'dovich (SZ) effect data.  To illustrate this, we compare
the scaling relations based on the estimated hydrostatic mass
($M_{500}^{\rm HSE}(<\!r_{500}^{\rm est})$) derived from the mock
\emph{Chandra} analysis of simulated clusters to the \emph{Chandra}
measurements in the lower panels of Figure~\ref{fig:xscale}.  Note
that we account for additional biases in the estimated cluster masses
arising from a bias in the estimation of a cluster virial radius by
measuring the cluster mass within the $r_{500}^{\rm est}$ estimated
from the hydrostatic analysis \citep[see also][for more details and
discussions]{nagai_etal07}.  Similarly, the gas mass ($M_{g,500}$) is
also computed at $r_{500}^{\rm true}$ and $r_{500}^{\rm est}$ in the
upper and lower panels.\footnote{Note that $\Tx$ is computed within
$r_{500}^{\rm true}$ in both panels; however, correcting for the bias
in $r_{500}$ has a negligible ($\lesssim 1\%$) effect on the $\Tx$
estimate.}  In Table~\ref{tab:plfit}, we summarize results for the
relaxed clusters at z=0, relevant for comparison with observations
considered here.  These analyses show that the simulation results are
in much better agreement with observations when using the hydrostatic
mass. The systematic offset in normalizations could thus be due to the
bias of total hydrostatic mass estimate due to turbulent motions of
the ICM.

In terms of scatter, the $\M500-\Tx$ relation exhibits the largest
scatter of $\sim 20\%$ in $\M500$ around the mean relation, most of
which is due to unrelaxed clusters.  The unrelaxed clusters also have
temperatures biased low for a given mass, likely because the mass of
the system has already increased but only a fraction of the kinetic
energy of merging systems is converted into the thermal energy of gas,
due to incomplete relaxation during mergers \citep{mathiesen_etal01}.
Unfortunately, we cannot compare the scatter directly to the
\emph{Chandra} results because for real clusters, the scatter is
dominated by the measurement uncertainties and the intrinsic scatter
\citep[see discussion in][]{vikhlinin_etal06}. The slope and redshift
evolution of the $\M500-\Tx$ relations are quite close to the simple
self-similar expectation.  The $\M500-\Mg$ relation has a somewhat
smaller scatter ($\approx 11\%$) around the best-fit power-law
relation than the $\M500-\Tx$, but its slope is significantly
different from the self-similar prediction for the $\M500-\Mg$
relation due to the trend of gas fraction with cluster mass present
for both the simulated clusters in our sample
\citep[see][]{kravtsov_etal05} and for the observed clusters
\citep{vikhlinin_etal06}. In all of the scaling relations considered
here, the use of $M^{\rm HSE}_{500}$, instead of $M^{\rm true}_{500}$,
modifies the scatter by a few percent for the relaxed clusters.

The $\M500-Y_X$ relation shows the scatter of only $\approx 7\%$,
making it by far the tightest of all the scaling relations.  Note that
this value of scatter includes clusters at both low and high-redshifts
and both relaxed and unrelaxed systems. The tightness of the
$\M500-\Yx$ relation and simple evolution are due to a fortunate
cancellation of opposite trends in gas mass and temperature
\citep[see][]{kravtsov_etal06}.  The slope and redshift evolution of
normalization for the $\M500-Y_X$ relations are well-described by the
simple self-similar model. 

\begin{deluxetable}{ccccc}
\tablecaption{Best-fit parameters and scatter in the mass vs.{} proxy
  relations, $M_{500}=CX^{\alpha}$, for relaxed clusters at $z=0$ and
  0.6.\label{tab:plfit}} 
\tablehead{
\nls
\multicolumn{1}{c}{relation\tablenotemark{a}}&
\multicolumn{1}{c}{quantity\tablenotemark{b}}&
\multicolumn{1}{c}{$M_{\rm 500}^{\rm SIM}$(True)} &
\multicolumn{1}{c}{$M_{\rm 500}^{\rm SIM}$(HSE)} &
\multicolumn{1}{c}{$M_{\rm 500}$(\emph{Chandra})} 
\nls
}
\startdata
\nls
$M_{\rm 500}-\Tx$ \dotfill & $\log_{10}C$  & $14.21\pm 0.010$ & $14.10\pm 0.008$ & $14.11\pm 0.035$ \\
             & $\alpha$ & $1.515\pm 0.052$  & $1.623\pm 0.027$ & $1.489\pm 0.093$ \\
             & scatter  & 0.136 & 0.117 & \nodata \\
             & $\log_{10}C_{\rm ss}$      & $14.21\pm 0.009$ & $14.10\pm 0.009$ & $14.10\pm 0.014$ \\
\nlss
$M_{\rm 500}-\Mg$ \dotfill & $\log_{10}C$  & $14.12\pm 0.008$ & $14.05\pm 0.011$ & $14.14\pm 0.044$ \\
             & $\alpha$ & $0.894\pm 0.023$  & $0.932\pm 0.033$ & $0.811\pm 0.067$ \\
             & scatter  & 0.114 & 0.153 & \nodata \\
             & $\log_{10}C_{\rm ss}$     & $14.12\pm 0.010$ & $14.05\pm 0.011$ & $14.07\pm 0.022$ \\
\nlss
$M_{\rm 500}-\Yx$ \dotfill  & $\log_{10}C$  & $14.06\pm 0.004$ & $13.97\pm 0.007$ & $14.04\pm 0.047$ \\
             & $\alpha$ & $0.568\pm 0.006$  & $0.596\pm 0.010$ & $0.526\pm 0.038$ \\
             & scatter  & 0.053 & 0.087 & \nodata \\
             & $\log_{10}C_{\rm ss}$     & $14.05\pm 0.005$ & $13.97\pm 0.006$ & $13.98\pm 0.017$ 
\nls
\enddata
\tablenotetext{a}{Power law fits were performed for relaxed
  clusters of our cluster sample at $z=0$ and $0.6$. In addition to the
  fits in which both normalization and slope of the power-law
  relations were fit simultaneously, we provide the best fit
  normalizations, $C_{\rm ss}$, for each relation when fit with the
  slopes fixed to their self-similar values: $1.5$, $1.0$, and $0.6$
  for the $\M500-\Tx$, $\M500-\Mg$, and $\M500-\Yx$ relations,
  respectively.} 
\tablenotetext{b}{For each observable $X$ ($=\Tx$, $\Mg$, $\Yx$),
  we fit power law relation of the form $\M500=C(X/X_0)^{\alpha}$,
  with $X_0=3.0$~keV, $2\times 10^{13}\,M_{\odot}$, and $4\times
  10^{13}$~keV~$M_\odot$, for $\Tx$, $\Mg$, $\Yx$, respectively. Note
  that $\M500$ is in units of $(h^{-1} M_{\odot})$.} 
\end{deluxetable}

\section{Discussion and conclusions}
\label{sec:discussion}

We presented analyses of the simulated cluster sample of 16 clusters
spanning a representative mass range ($5\times 10^{13}-2\times
10^{15}\,h^{-1}\,M_{\odot}$) and modeled using the shock-capturing
Eulerian adaptive mesh refinement $N$-body+gasdynamics ART code in the
$\Lambda$CDM cosmology.  These simulations achieve high spatial
resolution and include various physical processes of galaxy formation,
including radiative cooling, star formation and other processes
accompanying galaxy formation.  We study the effects of gas cooling
and star formation on the ICM properties by comparing two sets of
simulations performed with and without these processes included.  The
results of simulations with dissipation are compared to recent
\emph{Chandra} X-ray observations of nearby, relaxed clusters
\citep{vikhlinin_etal05c,vikhlinin_etal06}.

We show that gas cooling and star formation modify both the
normalization and the shape of the gas density, temperature, entropy,
and pressure profiles. As the lowest-entropy gas cools and condenses
out of the hot phase in the cluster progenitors, the gas density in
their inner regions is lowered and entropy is increased
\citep{bryan00,voit_bryan01}.  The effects have strong radial
dependence and are the strongest in the inner regions $r\lesssim
0.1r_{500}$.  At these inner radii simulation profiles do not match
the observations.  On the other hand, at $r\gtrsim 0.1r_{500}$ the
profiles in the CSF simulations and observations agree quite well,
while profiles in the non-radiative runs disagree with observations at
all radii within $r_{500}$. 

In particular, the simulations with cooling can explain the observed
high levels of entropy in observed clusters compared to the
non-radiative expectation pointed out previously
\citep{ponman_etal03,pratt_etal06}.  At $r\gtrsim r_{2500}$ the
cluster profiles are approximately self-similar within current
statistical error bars, while there is an indication that their
best-fit slopes are slightly shallower than the self-similar value.
Moreover, the slope and normalization of the entropy scaling relations
in the simulated clusters are in good agreement with \emph{Chandra}
observations at $r_{1000}$ and $r_{500}$, while the observed relations
exhibits deviations from the self-similarity at $r\lesssim r_{2500}$.
Note also that the results of \emph{Chandra} and \emph{XMM-Newton}
measurements agree quite well within $r\lesssim r_{2500}$, but the
significant disagreement is seen at $r_{1000}$. The statistical
significance of the difference between the \emph{Chandra} and
\emph{XMM-Newton} slopes is $3.4\sigma$ for the $K-M$ and $2.4\sigma$
for the $K-T$ relations. The difference is critical for theoretical
interpretation and implications, as the \emph{XMM-Newton} scaling was
billed as a major evidence for deviations from self-similar scalings
at large radii (indeed at $r_{1000}$, where \emph{XMM-Newton}
measurements is several sigma away from the slope of 1.0).
\emph{Chandra} results show that deviations at larger radii, if they
exist, are much smaller.  Despite the deviations from self-similarity
in the ICM entropies, we show that pressure profiles in particular,
show a remarkable degree of self-similarity and exhibit very small
scatter.

We also present comparisons of scaling relations of cluster X-ray
observables with total cluster mass in the simulations with cooling
and recent deep \emph{Chandra} observations. Specifically, we compare
correlation of the spectral X-ray temperature, ICM gas mass ($\Mgas$),
and the X-ray equivalent of integrated pressure ($Y_X\equiv
\Mgas\Tx$).  In these comparisons X-ray observables for the simulated
clusters are derived from mock \emph{Chandra} analysis using procedure
essentially identical to those used in real data analysis. 

The slope and normalization of the $M-\Tx$ and $M-\Yx$ relations in
simulations and \emph{Chandra} observations are in good agreement, and
they are consistent with the simple self-similar expectation.  In terms
of scatter, the $\M500-Y_X$ relation shows scatter of only $\approx
7\%$, making it by far the tightest of all the scaling relations.  Note
that this value of scatter includes clusters at both low and high
redshifts and both relaxed and unrelaxed systems.  The $\M500-\Tx$
relation, on the other hand, exhibits the largest scatter of $\sim 20\%$
scatter in $\M500$ around the mean relation, most of which is due to
unrelaxed clusters.  The unrelaxed clusters also have temperatures
biased low for a given mass, likely because the mass of the system has
already increased but only a fraction of the kinetic energy of merging
systems is converted into the thermal energy of gas, due to incomplete
relaxation \citep{mathiesen_etal01} during mergers. 

Moreover, these comparisons show that the normalizations of the
scaling relations of relaxed clusters in simulations and observations
agree at a level of about $\approx 10\%-20\%$.  This is a considerable
improvement, given that significant disagreement existed just several
years ago \citep[see][]{finoguenov_etal01,seljak02,pierpaoli_etal03}.
The residual systematic offset in the normalization is likely caused
by non-thermal pressure support from subsonic turbulent gas motions
\citep[][E. Lau et al. 2007, in
preparation]{evrard_etal96,rasia_etal04,rasia_etal06,faltenbacher_etal05,dolag_etal05}.
This contribution is approximately independent of cluster mass
\citep[][E. Lau et al. 2007, in preparation]{vazza_etal06} and is not
accounted for in X-ray hydrostatic mass estimates. For example, when
we repeat the comparison of scaling relations using hydrostatic mass
estimates for the observed clusters, we find excellent agreement in
normalizations, demonstrating explicitly that there is a systematic
$\approx 10\%-20\%$ offset between hydrostatic mass estimate and the
true mass in simulated clusters.

Part of the non-thermal pressure support may also be contributed by
cosmic-rays and magnetic fields.  In practice, it may be difficult to
distinguish between different sources of non-thermal pressure. A
possible test is their radial dependence. Turbulent motions, for
example, are in general smaller at smaller radii and the turbulent
pressure gradient is correspondingly smaller. In the case of turbulent
motions, we can therefore expect that the bias in the total mass
estimate should decrease at smaller radii.  This may not be the case
for some other sources of non-thermal pressure, although recent models
of cosmic-rays contribution to the total pressure show a qualitatively
similar radial dependence as the turbulent pressure
\citep{pfrommer_etal07}.

The much improved agreement of the scaling relations and, especially,
normalization and shape of the gas profiles between simulations with
cooling and star formation and observations show that inclusion of
galaxy formation in cluster simulations results in more realistic
modeling of the hot ICM. This may indicate that gravitational dynamics
and the basic cooling of the hot gas accompanying galaxy formation are
the dominant processes determining thermodynamics of the ICM outside
the cluster cores, while other processes, such as feedback, thermal
conduction, viscosity, and cosmic-rays, are playing only a minor role
for a large fraction of the ICM mass.

Note, however, that the agreement between our simulations and
observations is achieved by condensation of a significant fraction of
hot gas into cold dense phase, which is subsequently converted into
stars. Thus, simulated clusters have $\approx 40\%$ of their baryons
within $\r500$ in stellar form at $z=0$, while the rest of the baryons
are in the hot phase. Although the low hot gas mass fractions
($\approx 60\%-70\%$ of the universal value) are consistent with
observations \citep{vikhlinin_etal06,mccarthy_etal07}, the high
stellar fractions are not.  Note, however, that the reduced stellar
fraction with more efficient stellar or AGN energy feedback generally
results in the profiles that are in between those of the CSF and
non-radiative runs and disrupts the good agreement between models and
data. 

Observational estimates of the stellar mass fractions in groups and
clusters range from $\approx 5\%-10\%$ \citep{eke_etal05} to $\approx
15\%-20\%$ \citep{lin_etal03,gonzalez_etal07} of the universal baryon
fraction or at least a factor of 2-3 lower than the fractions found in
our simulations. This is a well-known discrepancy often referred to as
the ``overcooling problem''. Our results show that the X-ray and
optical observations appear to give seemingly contradicting
constraints. The low observed stellar fractions imply existence of an
efficient mechanism suppressing star formation in real clusters, while
observed properties of hot ICM are not consistent with small amounts
of cooling (i.e., predictions close to the non-radiative limit). At
present it is not clear how these two observational constraints can be
reconciled.

Our tests indicate that profiles and average quantities (i.e., gas
fractions) derived from analyses of modern X-ray data are robust and
do not suffer any obvious biases \citep{nagai_etal07}.  On the other
hand, there are certain systematic uncertainties in estimates of
stellar mass from optical observations related both to possible
low-surface stellar component missed in shallow observations
\citep[e.g.,][]{gonzalez_etal05,gonzalez_etal07,lauer_etal07,seigar_etal07}
and to the uncertainties in the stellar population modeling of the
observed photometry. It is unlikely, however, that any single
uncertainty is large enough to account for the entire factor of 2-3
discrepancy between stellar fractions in simulations and observations.
We note also that stellar fraction predicted by simulations depends on
the implementation of the feedback processes in simulations
\citep[e.g.,][]{borgani_etal06}. However, the current implementations
of the feedback processes efficient in significantly suppressing
stellar fraction are essentially ad hoc and it is uncertain whether
the feedback is actually as efficient in practice.

The progress in our understanding of these issues should come from
detailed convergence studies and comparisons of simulation results
done using different numerical codes, further comparisons of
simulations with deep X-ray observations, deeper observations, and
thorough analysis of uncertainties in the optical estimates of cluster
stellar masses.

\acknowledgements We would like to thank Monique Arnaud, Trevor
Ponman, Gabriel Pratt, and Alastair Sanderson for providing their
observational data points.  We also thank Monique Arnaud, Greg Bryan,
Gus Evrard, Gabriel Pratt, Alastair Sanderson, Riccardo Valdarnini,
and Mark Voit for useful comments on the manuscript. D. N. is supported
by the Sherman Fairchild Postdoctoral Fellowship at Caltech.
A. V. K. is supported by the National Science Foundation (NSF) under
grants AST-0239759 and AST-0507666, by NASA through grant NAG5-13274,
and by the Kavli Institute for Cosmological Physics at the University
of Chicago.  A. V. is supported by the NASA contract NAS8-39073 and
\emph{Chandra} General Observer grant GO5-6121A. The cosmological
simulations used in this study were performed on the IBM RS/6000 SP4
system ({\tt copper}) at the National Center for Supercomputing
Applications (NCSA).

\appendix
\section{Analytic pressure model}
\label{sec:pressure_model}

The Sunyaev-Zel'dovich (SZ) effect is a direct probe of thermal energy
content of the Universe and provides a unique and powerful probe of
the structure formation and cosmology in the near future. The SZ
observations are probing the integrated pressure of the ICM. Accurate
analytic parameterizations of the ICM pressure profiles can therefore
be useful for developing efficient cluster detection algorithm for
upcoming SZ cluster surveys, analysis and interpretation of SZ effect
observations, as well as theoretical modeling of cluster ICM. The fact
that the self-similarity is best preserved for the pressure profiles
and their low cluster-to-cluster scatter (see
\S~\ref{sec:profiles_sim}) provides further motivation for the use of
accurate pressure profile parameterizations.

\begin{figure*}[t]  
  \vspace{0.0cm}
  \centerline{ \epsfysize=4.5truein \epsffile{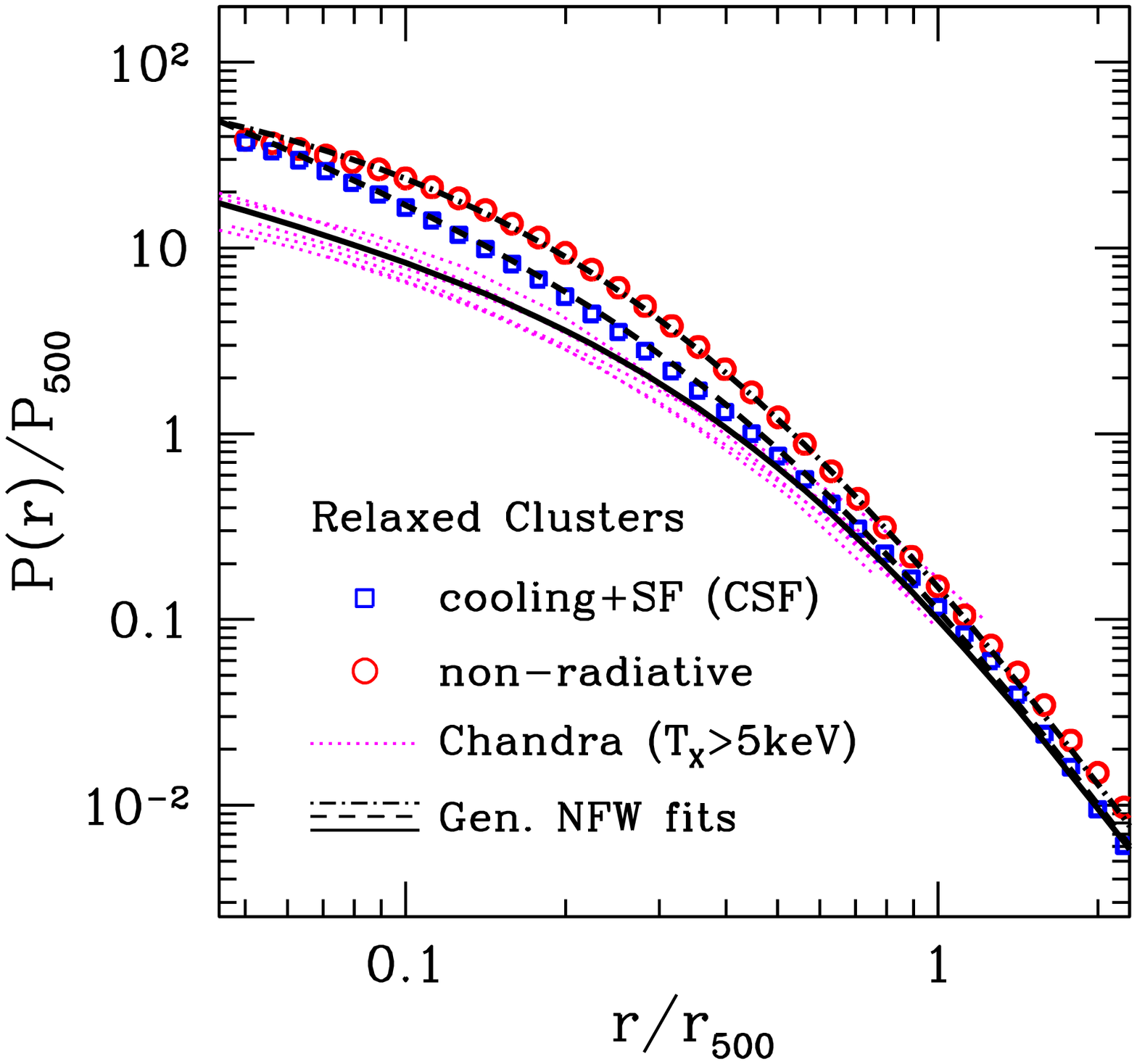} }
  \vspace{-0.5cm}  
  \caption{Generalized NFW fits to the pressure profiles of relaxed
  clusters in simulations and \emph{Chandra} X-ray
  observations. \emph{Open circles} and \emph{squares} show the mean
  profiles in the CSF and non-radiative simulations, respectively.
  \emph{Thin} dotted shows \emph{Chandra} X-ray clusters with $T_{\rm
  X}>5$~keV.  \emph{Thick} lines show the best-fit generalized NFW
  model to simulation and observed profiles (see Table~\ref{tab:pfits}
  for the best-fit parameters). }
\label{fig:pfits}
\end{figure*}

Here we present a simple analytic model of the pressure profile that
closely matches the observed profiles of the \emph{Chandra} X-ray clusters
and results of numerical simulations in their outskirts.  Since the
gas pressure distribution is primarily determined by the
gravitationally dominant dark matter component, we parameterize the
pressure profile using the generalized NFW model,
\begin{equation}
\frac{P(r)}{P_{500}} = \frac{P_0}{ x^{\gamma} ( 1 + x^{\alpha}
  )^{(\beta-\gamma) / \alpha}}
\label{eq:pmodel}
\end{equation}
where $x \equiv r/r_s$, $r_s=r_{500}/c_{500}$, $P_{500}$ is given by
equation~\ref{eq:p500}, and $(\alpha, \beta, \gamma)$ are the slopes
at $r\sim r_s$, $r\gg r_s$, and $r\ll r_s$, respectively.  We find
that a model with $P_0$=3.3, $c_{500}\equiv r_{500}/r_s$=1.8, and
$(\alpha, \beta, \gamma)=(1.3, 4.3, 0.7)$ provides a good description
of the pressure profiles of the high-$\Tx$ \emph{Chandra} clusters
within the observed range ($r\lesssim r_{500}$) as well as the
profiles of simulated clusters in
$0.5<r/r_{500}<2.0$. Figure~\ref{fig:pfits} shows generalized NFW fits
to the pressure profiles of relaxed clusters in simulations and
\emph{Chandra} observations. In the outskirts, we set the slope to be
$\gamma = 4.3$, which is the average best-fit values for both
non-radiative and CSF simulations. For the \emph{Chandra} clusters
with $\Tx>5$~keV, the inner slopes of the pressure profile are
$\approx 0.7$.  The inner slopes appear to be shallower for the
lower-$\Tx$ systems, but they also show much larger cluster-to-cluster
variation.  Table~\ref{tab:pfits} summarizes the best-fit model
parameters for observed and simulated clusters.  For the relaxed
systems, the same set of parameters with a different inner slope
$\gamma=1.1$ produces the pressure profile of the CSF run, while that
of the non-radiative run requires a very different set of parameters.
The pressure profiles of the unrelaxed systems are generally less
concentrated (smaller value of $c_{500}$) with slightly different
inner and outer slopes.

\begin{deluxetable}{ccccc}
  \tablecaption{Best-fit parameters of the pressure
    profile.\label{tab:pfits}} \tablehead{ \colhead{Runs} &
    \colhead{Sample} & \colhead{$P_0$} & \colhead{$c_{500}$} &
    \colhead{$(\alpha, \beta, \gamma)$} } \startdata
  Chandra Obs.  & Relaxed   & $3.3$  &  $1.8$  & $(1.3,4.3,0.7)$ \\
  Cooling+SF    & Relaxed   & $3.3$  &  $1.8$  & $(1.3,4.3,1.1)$ \\
  Cooling+SF    & Unrelaxed & $2.0$  &  $1.5$  & $(1.4,4.3,1.0)$ \\
  Non-radiative & Relaxed   & $38.0$ &  $3.0$  & $(1.1,4.3,0.3)$ \\
  Non-radiative & Unrelaxed & $3.0$  &  $1.5$  & $(1.4,4.3,0.9)$
\enddata
\end{deluxetable}

\end{document}